\documentclass[longauth]{aa}

\usepackage{graphicx}
\usepackage{natbib}
\usepackage{scalerel}

\usepackage[table]{xcolor}

\usepackage{txfonts}
\usepackage[pdfencoding=auto,psdextra]{hyperref}
\hypersetup{
    colorlinks=true,
    linkcolor=blue,
    filecolor=magenta,      
    urlcolor=blue,
    citecolor=blue
}
\makeatletter
\renewcommand*\aa@pageof{, page \thepage{} of \pageref*{LastPage}}
\makeatother

\usepackage[utf8]{inputenc}

\usepackage[switch, modulo]{lineno}
              
\renewcommand{\linenumbers}[0]{}

\usepackage{euclid}

\newcommand{\Gaia}{\textit{Gaia}}
\newcommand{\twomass}{2MASS}
\newcommand{\neff}{\ensuremath{n^{\mathrm{spec}}_{\mathrm{eff}}}}
\newcommand{\sqdeg}{\ensuremath{\,\mathrm{deg}^2}} % aggiunge anche uno spazio fine prima dell’unità

\usepackage{verbatim}

\begin{document}
%
% Put the title and authors of your (Standard Project) paper here
%

%\title{\Euclid\/: A two-point correlation approach to diagnosing star-related systematics in the \Euclid spectroscopic survey}
\title{\Euclid}
\subtitle{A two-point correlation approach to diagnosing star-related systematics in the spectroscopic survey}

%%%% Version Tuesday 1st of September 2026 01:19:00 PM UT
%%%% Assumes the new A&A style file from Oct 2025 or later
%%%% Please do not edit the author list -- contact ECEB Bureau for changes
\newcommand{\orcid}[1]{} %% if already defined in aa.cls: comment, or use renewcommand			   
\author{Euclid Collaboration: I.~Risso\orcid{0000-0003-2525-7761}\thanks{\email{ilaria.risso@ge.infn.it}}\inst{\ref{aff1},\ref{aff2},\ref{aff3}}
\and B.~R.~Granett\orcid{0000-0003-2694-9284}\inst{\ref{aff3}}
\and E.~Branchini\orcid{0000-0002-0808-6908}\inst{\ref{aff1},\ref{aff2},\ref{aff3}}
\and A.~Veropalumbo\orcid{0000-0003-2387-1194}\inst{\ref{aff3},\ref{aff2},\ref{aff1}}
\and B.~Kubik\orcid{0009-0006-5823-4880}\inst{\ref{aff4}}
\and P.~Monaco\orcid{0000-0003-2083-7564}\inst{\ref{aff5},\ref{aff6},\ref{aff7},\ref{aff8}}
\and J.~Comparat\orcid{0000-0001-9200-1497}\inst{\ref{aff9}}
\and A.~Hall\orcid{0000-0002-3139-8651}\inst{\ref{aff10}}
\and N.~Aghanim\orcid{0000-0002-6688-8992}\inst{\ref{aff11}}
\and B.~Altieri\orcid{0000-0003-3936-0284}\inst{\ref{aff12}}
\and A.~Amara\inst{\ref{aff13}}
\and S.~Andreon\orcid{0000-0002-2041-8784}\inst{\ref{aff3}}
\and N.~Auricchio\orcid{0000-0003-4444-8651}\inst{\ref{aff14}}
\and C.~Baccigalupi\orcid{0000-0002-8211-1630}\inst{\ref{aff8},\ref{aff6},\ref{aff7},\ref{aff15}}
\and M.~Baldi\orcid{0000-0003-4145-1943}\inst{\ref{aff16},\ref{aff14},\ref{aff17}}
\and A.~Balestra\orcid{0000-0002-6967-261X}\inst{\ref{aff18}}
\and S.~Bardelli\orcid{0000-0002-8900-0298}\inst{\ref{aff14}}
\and P.~Battaglia\orcid{0000-0002-7337-5909}\inst{\ref{aff14}}
\and A.~Biviano\orcid{0000-0002-0857-0732}\inst{\ref{aff6},\ref{aff8}}
\and M.~Bolzonella\orcid{0000-0003-3278-4607}\inst{\ref{aff14}}
\and M.~Brescia\orcid{0000-0001-9506-5680}\inst{\ref{aff19},\ref{aff20}}
\and S.~Camera\orcid{0000-0003-3399-3574}\inst{\ref{aff21},\ref{aff22},\ref{aff23}}
\and G.~Ca\~nas-Herrera\orcid{0000-0003-2796-2149}\inst{\ref{aff24}}
\and V.~Capobianco\orcid{0000-0002-3309-7692}\inst{\ref{aff23}}
\and C.~Carbone\orcid{0000-0003-0125-3563}\inst{\ref{aff25}}
\and J.~Carretero\orcid{0000-0002-3130-0204}\inst{\ref{aff26},\ref{aff27}}
\and F.~J.~Castander\orcid{0000-0001-7316-4573}\inst{\ref{aff28},\ref{aff29}}
\and M.~Castellano\orcid{0000-0001-9875-8263}\inst{\ref{aff30}}
\and G.~Castignani\orcid{0000-0001-6831-0687}\inst{\ref{aff14}}
\and S.~Cavuoti\orcid{0000-0002-3787-4196}\inst{\ref{aff20},\ref{aff31}}
\and A.~Cimatti\inst{\ref{aff32}}
\and C.~Colodro-Conde\inst{\ref{aff33}}
\and G.~Congedo\orcid{0000-0003-2508-0046}\inst{\ref{aff10}}
\and L.~Conversi\orcid{0000-0002-6710-8476}\inst{\ref{aff34},\ref{aff12}}
\and Y.~Copin\orcid{0000-0002-5317-7518}\inst{\ref{aff4}}
\and F.~Courbin\orcid{0000-0003-0758-6510}\inst{\ref{aff35},\ref{aff36},\ref{aff37}}
\and H.~M.~Courtois\orcid{0000-0003-0509-1776}\inst{\ref{aff38}}
\and H.~Degaudenzi\orcid{0000-0002-5887-6799}\inst{\ref{aff39}}
\and S.~de~la~Torre\inst{\ref{aff40}}
\and G.~De~Lucia\orcid{0000-0002-6220-9104}\inst{\ref{aff6}}
\and H.~Dole\orcid{0000-0002-9767-3839}\inst{\ref{aff11}}
\and F.~Dubath\orcid{0000-0002-6533-2810}\inst{\ref{aff39}}
\and X.~Dupac\inst{\ref{aff12}}
\and S.~Dusini\orcid{0000-0002-1128-0664}\inst{\ref{aff41}}
\and S.~Escoffier\orcid{0000-0002-2847-7498}\inst{\ref{aff42}}
\and M.~Farina\orcid{0000-0002-3089-7846}\inst{\ref{aff43}}
\and R.~Farinelli\inst{\ref{aff14}}
\and S.~Ferriol\inst{\ref{aff4}}
\and F.~Finelli\orcid{0000-0002-6694-3269}\inst{\ref{aff14},\ref{aff44}}
\and N.~Fourmanoit\orcid{0009-0005-6816-6925}\inst{\ref{aff42}}
\and M.~Frailis\orcid{0000-0002-7400-2135}\inst{\ref{aff6}}
\and E.~Franceschi\orcid{0000-0002-0585-6591}\inst{\ref{aff14}}
\and M.~Fumana\orcid{0000-0001-6787-5950}\inst{\ref{aff25}}
\and L.~Gabarra\orcid{0000-0002-8486-8856}\inst{\ref{aff45}}
\and S.~Galeotta\orcid{0000-0002-3748-5115}\inst{\ref{aff6}}
\and K.~George\orcid{0000-0002-1734-8455}\inst{\ref{aff46}}
\and W.~Gillard\orcid{0000-0003-4744-9748}\inst{\ref{aff42}}
\and B.~Gillis\orcid{0000-0002-4478-1270}\inst{\ref{aff10}}
\and C.~Giocoli\orcid{0000-0002-9590-7961}\inst{\ref{aff14},\ref{aff17}}
\and J.~Gracia-Carpio\orcid{0000-0003-4689-3134}\inst{\ref{aff47}}
\and A.~Grazian\orcid{0000-0002-5688-0663}\inst{\ref{aff18}}
\and F.~Grupp\inst{\ref{aff47},\ref{aff48}}
\and L.~Guzzo\orcid{0000-0001-8264-5192}\inst{\ref{aff49},\ref{aff3},\ref{aff50}}
\and S.~V.~H.~Haugan\orcid{0000-0001-9648-7260}\inst{\ref{aff51}}
\and S.~Hemmati\orcid{0000-0003-2226-5395}\inst{\ref{aff52}}
\and W.~Holmes\orcid{0009-0007-8554-4646}\inst{\ref{aff53}}
\and F.~Hormuth\inst{\ref{aff54}}
\and A.~Hornstrup\orcid{0000-0002-3363-0936}\inst{\ref{aff55},\ref{aff56}}
\and K.~Jahnke\orcid{0000-0003-3804-2137}\inst{\ref{aff57}}
\and M.~Jhabvala\inst{\ref{aff58}}
\and B.~Joachimi\orcid{0000-0001-7494-1303}\inst{\ref{aff59}}
\and S.~Kermiche\orcid{0000-0002-0302-5735}\inst{\ref{aff42}}
\and A.~Kiessling\orcid{0000-0002-2590-1273}\inst{\ref{aff53}}
\and M.~K\"ummel\orcid{0000-0003-2791-2117}\inst{\ref{aff48}}
\and M.~Kunz\orcid{0000-0002-3052-7394}\inst{\ref{aff60}}
\and H.~Kurki-Suonio\orcid{0000-0002-4618-3063}\inst{\ref{aff61},\ref{aff62}}
\and A.~M.~C.~Le~Brun\orcid{0000-0002-0936-4594}\inst{\ref{aff63}}
\and S.~Ligori\orcid{0000-0003-4172-4606}\inst{\ref{aff23}}
\and P.~B.~Lilje\orcid{0000-0003-4324-7794}\inst{\ref{aff51}}
\and V.~Lindholm\orcid{0000-0003-2317-5471}\inst{\ref{aff61},\ref{aff62}}
\and I.~Lloro\orcid{0000-0001-5966-1434}\inst{\ref{aff64}}
\and M.~Magliocchetti\orcid{0000-0001-9158-4838}\inst{\ref{aff43}}
\and G.~Mainetti\orcid{0000-0003-2384-2377}\inst{\ref{aff65}}
\and O.~Mansutti\orcid{0000-0001-5758-4658}\inst{\ref{aff6}}
\and O.~Marggraf\orcid{0000-0001-7242-3852}\inst{\ref{aff66}}
\and M.~Martinelli\orcid{0000-0002-6943-7732}\inst{\ref{aff30},\ref{aff67}}
\and N.~Martinet\orcid{0000-0003-2786-7790}\inst{\ref{aff40}}
\and F.~Marulli\orcid{0000-0002-8850-0303}\inst{\ref{aff68},\ref{aff14},\ref{aff17}}
\and R.~J.~Massey\orcid{0000-0002-6085-3780}\inst{\ref{aff69}}
\and N.~Mauri\orcid{0000-0001-8196-1548}\inst{\ref{aff32},\ref{aff17}}
\and E.~Medinaceli\orcid{0000-0002-4040-7783}\inst{\ref{aff14}}
\and M.~Melchior\inst{\ref{aff70}}
\and M.~Meneghetti\orcid{0000-0003-1225-7084}\inst{\ref{aff14},\ref{aff17}}
\and E.~Merlin\orcid{0000-0001-6870-8900}\inst{\ref{aff18}}
\and G.~Meylan\orcid{0000-0001-6503-0209}\inst{\ref{aff71}}
\and A.~Mora\orcid{0000-0002-1922-8529}\inst{\ref{aff72}}
\and M.~Moresco\orcid{0000-0002-7616-7136}\inst{\ref{aff68},\ref{aff14}}
\and C.~Moretti\orcid{0000-0003-3314-8936}\inst{\ref{aff6},\ref{aff8},\ref{aff7}}
\and L.~Moscardini\orcid{0000-0002-3473-6716}\inst{\ref{aff68},\ref{aff14},\ref{aff17}}
\and R.~Nakajima\orcid{0009-0009-1213-7040}\inst{\ref{aff66}}
\and C.~Neissner\orcid{0000-0001-8524-4968}\inst{\ref{aff73},\ref{aff27}}
\and S.-M.~Niemi\orcid{0009-0005-0247-0086}\inst{\ref{aff74}}
\and J.~W.~Nightingale\orcid{0000-0002-8987-7401}\inst{\ref{aff75}}
\and C.~Padilla\orcid{0000-0001-7951-0166}\inst{\ref{aff73}}
\and S.~Paltani\orcid{0000-0002-8108-9179}\inst{\ref{aff39}}
\and F.~Pasian\orcid{0000-0002-4869-3227}\inst{\ref{aff6}}
\and W.~J.~Percival\orcid{0000-0002-0644-5727}\inst{\ref{aff76},\ref{aff77},\ref{aff78}}
\and V.~Pettorino\orcid{0000-0002-4203-9320}\inst{\ref{aff74}}
\and A.~Pezzotta\orcid{0000-0003-0726-2268}\inst{\ref{aff3}}
\and S.~Pires\orcid{0000-0002-0249-2104}\inst{\ref{aff79}}
\and G.~Polenta\orcid{0000-0003-4067-9196}\inst{\ref{aff80}}
\and M.~Poncet\inst{\ref{aff81}}
\and L.~A.~Popa\inst{\ref{aff82}}
\and F.~Raison\orcid{0000-0002-7819-6918}\inst{\ref{aff47}}
\and A.~Renzi\orcid{0000-0001-9856-1970}\inst{\ref{aff83},\ref{aff41},\ref{aff14}}
\and J.~Rhodes\orcid{0000-0002-4485-8549}\inst{\ref{aff53}}
\and G.~Riccio\inst{\ref{aff20}}
\and F.~Rizzo\orcid{0000-0002-9407-585X}\inst{\ref{aff6}}
\and E.~Romelli\orcid{0000-0003-3069-9222}\inst{\ref{aff6}}
\and M.~Roncarelli\orcid{0000-0001-9587-7822}\inst{\ref{aff14}}
\and B.~Rusholme\orcid{0000-0001-7648-4142}\inst{\ref{aff52}}
\and R.~Saglia\orcid{0000-0003-0378-7032}\inst{\ref{aff48},\ref{aff47}}
\and Z.~Sakr\orcid{0000-0002-4823-3757}\inst{\ref{aff84},\ref{aff85},\ref{aff86}}
\and A.~G.~S\'anchez\orcid{0000-0003-1198-831X}\inst{\ref{aff47}}
\and D.~Sapone\orcid{0000-0001-7089-4503}\inst{\ref{aff87}}
\and B.~Sartoris\orcid{0000-0003-1337-5269}\inst{\ref{aff47},\ref{aff6}}
\and P.~Schneider\orcid{0000-0001-8561-2679}\inst{\ref{aff66}}
\and T.~Schrabback\orcid{0000-0002-6987-7834}\inst{\ref{aff88}}
\and M.~Scodeggio\inst{\ref{aff25}}
\and A.~Secroun\orcid{0000-0003-0505-3710}\inst{\ref{aff42}}
\and E.~Sihvola\orcid{0000-0003-1804-7715}\inst{\ref{aff89}}
\and C.~Sirignano\orcid{0000-0002-0995-7146}\inst{\ref{aff83},\ref{aff41}}
\and G.~Sirri\orcid{0000-0003-2626-2853}\inst{\ref{aff17}}
\and L.~Stanco\orcid{0000-0002-9706-5104}\inst{\ref{aff41}}
\and P.~Tallada-Cresp\'{i}\orcid{0000-0002-1336-8328}\inst{\ref{aff26},\ref{aff27}}
\and A.~N.~Taylor\inst{\ref{aff10}}
\and I.~Tereno\orcid{0000-0002-4537-6218}\inst{\ref{aff90},\ref{aff91}}
\and N.~Tessore\orcid{0000-0002-9696-7931}\inst{\ref{aff92}}
\and S.~Toft\orcid{0000-0003-3631-7176}\inst{\ref{aff93},\ref{aff94}}
\and R.~Toledo-Moreo\orcid{0000-0002-2997-4859}\inst{\ref{aff95},\ref{aff96}}
\and F.~Torradeflot\orcid{0000-0003-1160-1517}\inst{\ref{aff27},\ref{aff26}}
\and I.~Tutusaus\orcid{0000-0002-3199-0399}\inst{\ref{aff28},\ref{aff29},\ref{aff85}}
\and J.~Valiviita\orcid{0000-0001-6225-3693}\inst{\ref{aff61},\ref{aff62}}
\and T.~Vassallo\orcid{0000-0001-6512-6358}\inst{\ref{aff6},\ref{aff46}}
\and Y.~Wang\orcid{0000-0002-4749-2984}\inst{\ref{aff52}}
\and J.~Weller\orcid{0000-0002-8282-2010}\inst{\ref{aff48},\ref{aff47}}
\and G.~Zamorani\orcid{0000-0002-2318-301X}\inst{\ref{aff14}}
\and F.~M.~Zerbi\orcid{0000-0002-9996-973X}\inst{\ref{aff3}}
\and E.~Zucca\orcid{0000-0002-5845-8132}\inst{\ref{aff14}}
\and M.~Ballardini\orcid{0000-0003-4481-3559}\inst{\ref{aff97},\ref{aff98},\ref{aff14}}
\and E.~Bozzo\orcid{0000-0002-8201-1525}\inst{\ref{aff39}}
\and C.~Burigana\orcid{0000-0002-3005-5796}\inst{\ref{aff99},\ref{aff44}}
\and R.~Cabanac\orcid{0000-0001-6679-2600}\inst{\ref{aff85}}
\and M.~Calabrese\orcid{0000-0002-2637-2422}\inst{\ref{aff100},\ref{aff25}}
\and A.~Cappi\inst{\ref{aff101},\ref{aff14}}
\and F.~Caro\orcid{0009-0003-1053-0507}\inst{\ref{aff30}}
\and T.~Castro\orcid{0000-0002-6292-3228}\inst{\ref{aff102},\ref{aff6},\ref{aff8}}
\and J.~A.~Escartin~Vigo\inst{\ref{aff47}}
\and J.~Garc\'ia-Bellido\orcid{0000-0002-9370-8360}\inst{\ref{aff84}}
\and T.~Gasparetto\orcid{0000-0002-7913-4866}\inst{\ref{aff30}}
\and J.~Macias-Perez\orcid{0000-0002-5385-2763}\inst{\ref{aff9}}
\and R.~Maoli\orcid{0000-0002-6065-3025}\inst{\ref{aff103},\ref{aff30}}
\and R.~B.~Metcalf\orcid{0000-0003-3167-2574}\inst{\ref{aff68},\ref{aff14}}
\and A.~A.~Nucita\inst{\ref{aff104},\ref{aff105},\ref{aff106}}
\and M.~P\"ontinen\orcid{0000-0001-5442-2530}\inst{\ref{aff61}}
\and E.~Sarpa\orcid{0000-0002-1256-655X}\inst{\ref{aff6}}
\and V.~Scottez\orcid{0009-0008-3864-940X}\inst{\ref{aff107},\ref{aff108}}
\and M.~Sereno\orcid{0000-0003-0302-0325}\inst{\ref{aff14},\ref{aff17}}
\and M.~Tenti\orcid{0000-0002-4254-5901}\inst{\ref{aff17}}
\and M.~Tucci\inst{\ref{aff39}}
\and M.~Viel\orcid{0000-0002-2642-5707}\inst{\ref{aff8},\ref{aff6},\ref{aff15},\ref{aff7},\ref{aff109}}
\and M.~Wiesmann\orcid{0009-0000-8199-5860}\inst{\ref{aff51}}
\and J.~A.~Acevedo~Barroso\orcid{0000-0002-9654-1711}\inst{\ref{aff53}}
\and Y.~Akrami\orcid{0000-0002-2407-7956}\inst{\ref{aff84},\ref{aff110}}
\and I.~T.~Andika\orcid{0000-0001-6102-9526}\inst{\ref{aff48}}
\and S.~Anselmi\orcid{0000-0002-3579-9583}\inst{\ref{aff41},\ref{aff83},\ref{aff111}}
\and M.~Archidiacono\orcid{0000-0003-4952-9012}\inst{\ref{aff49},\ref{aff50}}
\and G.~Aric\`o\orcid{0000-0002-2802-2928}\inst{\ref{aff17}}
\and F.~Atrio-Barandela\orcid{0000-0002-2130-2513}\inst{\ref{aff112}}
\and M.~Baes\orcid{0000-0002-3930-2757}\inst{\ref{aff113}}
\and L.~Bazzanini\orcid{0000-0003-0727-0137}\inst{\ref{aff97},\ref{aff14}}
\and J.~Bel\orcid{0009-0006-7837-1866}\inst{\ref{aff114}}
\and D.~Bertacca\orcid{0000-0002-2490-7139}\inst{\ref{aff83},\ref{aff18},\ref{aff41}}
\and M.~Bethermin\orcid{0000-0002-3915-2015}\inst{\ref{aff115}}
\and F.~Beutler\orcid{0000-0003-0467-5438}\inst{\ref{aff10}}
\and L.~Blot\orcid{0000-0002-9622-7167}\inst{\ref{aff116},\ref{aff63}}
\and M.~Bonici\orcid{0000-0002-8430-126X}\inst{\ref{aff76},\ref{aff25}}
\and M.~L.~Brown\orcid{0000-0002-0370-8077}\inst{\ref{aff117}}
\and S.~Bruton\orcid{0000-0002-6503-5218}\inst{\ref{aff118}}
\and B.~Camacho~Quevedo\orcid{0000-0002-8789-4232}\inst{\ref{aff8},\ref{aff15},\ref{aff6}}
\and P.~Carrilho\orcid{0000-0003-1339-0194}\inst{\ref{aff119}}
\and C.~S.~Carvalho\inst{\ref{aff91}}
\and F.~Cogato\orcid{0000-0003-4632-6113}\inst{\ref{aff68},\ref{aff14}}
\and T.~E.~Collett\orcid{0000-0001-5564-3140}\inst{\ref{aff120}}
\and S.~Conseil\orcid{0000-0002-3657-4191}\inst{\ref{aff4}}
\and S.~Contarini\orcid{0000-0002-9843-723X}\inst{\ref{aff47}}
\and A.~R.~Cooray\orcid{0000-0002-3892-0190}\inst{\ref{aff121}}
\and P.~Corcho-Caballero\orcid{0000-0001-6327-7080}\inst{\ref{aff122}}
\and B.~Csizi\orcid{0000-0003-3227-6581}\inst{\ref{aff88}}
\and O.~Cucciati\orcid{0000-0002-9336-7551}\inst{\ref{aff14}}
\and S.~Davini\orcid{0000-0003-3269-1718}\inst{\ref{aff2}}
\and T.~de~Boer\orcid{0000-0001-5486-2747}\inst{\ref{aff123}}
\and F.~De~Paolis\orcid{0000-0001-6460-7563}\inst{\ref{aff104},\ref{aff105},\ref{aff106}}
\and G.~Desprez\orcid{0000-0001-8325-1742}\inst{\ref{aff122}}
\and A.~D\'iaz-S\'anchez\orcid{0000-0003-0748-4768}\inst{\ref{aff124}}
\and S.~Di~Domizio\orcid{0000-0003-2863-5895}\inst{\ref{aff1},\ref{aff2}}
\and J.~M.~Diego\orcid{0000-0001-9065-3926}\inst{\ref{aff125}}
\and V.~Duret\orcid{0009-0009-0383-4960}\inst{\ref{aff42}}
\and A.~Enia\orcid{0000-0002-0200-2857}\inst{\ref{aff14}}
\and Y.~Fang\orcid{0000-0002-0334-6950}\inst{\ref{aff48}}
\and A.~Farina\orcid{0009-0000-3420-929X}\inst{\ref{aff3},\ref{aff2}}
\and A.~Finoguenov\orcid{0000-0002-4606-5403}\inst{\ref{aff61}}
\and A.~Franco\orcid{0000-0002-4761-366X}\inst{\ref{aff104},\ref{aff105},\ref{aff106}}
\and Y.~Fu\orcid{0000-0002-0759-0504}\inst{\ref{aff24},\ref{aff122}}
\and K.~Ganga\orcid{0000-0001-8159-8208}\inst{\ref{aff126}}
\and E.~Gaztanaga\orcid{0000-0001-9632-0815}\inst{\ref{aff28},\ref{aff29},\ref{aff120}}
\and Z.~Ghaffari\orcid{0000-0002-6467-8078}\inst{\ref{aff6},\ref{aff8}}
\and F.~Giacomini\orcid{0000-0002-3129-2814}\inst{\ref{aff17}}
\and F.~Gianotti\orcid{0000-0003-4666-119X}\inst{\ref{aff14}}
\and G.~Gozaliasl\orcid{0000-0002-0236-919X}\inst{\ref{aff127},\ref{aff61}}
\and A.~Gruppuso\orcid{0000-0001-9272-5292}\inst{\ref{aff14},\ref{aff17}}
\and M.~Guidi\orcid{0000-0001-9408-1101}\inst{\ref{aff16},\ref{aff14}}
\and C.~M.~Gutierrez\orcid{0000-0001-7854-783X}\inst{\ref{aff33},\ref{aff128}}
\and H.~Hildebrandt\orcid{0000-0002-9814-3338}\inst{\ref{aff129}}
\and J.~Hjorth\orcid{0000-0002-4571-2306}\inst{\ref{aff130}}
\and M.~Jauzac\orcid{0000-0003-1974-8732}\inst{\ref{aff131},\ref{aff69},\ref{aff132},\ref{aff133},\ref{aff85}}
\and J.~J.~E.~Kajava\orcid{0000-0002-3010-8333}\inst{\ref{aff134},\ref{aff135},\ref{aff136}}
\and Y.~Kang\orcid{0009-0000-8588-7250}\inst{\ref{aff88}}
\and V.~Kansal\orcid{0000-0002-4008-6078}\inst{\ref{aff137},\ref{aff138}}
\and D.~Karagiannis\orcid{0000-0002-4927-0816}\inst{\ref{aff97},\ref{aff139}}
\and J.~Kim\orcid{0000-0003-2776-2761}\inst{\ref{aff45}}
\and C.~C.~Kirkpatrick\inst{\ref{aff89}}
\and A.~Kov\'acs\orcid{0000-0002-5825-579X}\inst{\ref{aff140},\ref{aff141}}
\and I.~Kova{\v{c}}i{\'{c}}\orcid{0000-0001-6751-3263}\inst{\ref{aff113}}
\and K.~Koyama\orcid{0000-0001-6727-6915}\inst{\ref{aff120}}
\and S.~Kruk\orcid{0000-0001-8010-8879}\inst{\ref{aff12}}
\and M.~C.~Lam\orcid{0000-0002-9347-2298}\inst{\ref{aff10}}
\and F.~Leclercq\orcid{0000-0002-9339-1404}\inst{\ref{aff142}}
\and L.~Legrand\orcid{0000-0003-0610-5252}\inst{\ref{aff143},\ref{aff11}}
\and M.~Lembo\orcid{0000-0002-5271-5070}\inst{\ref{aff142}}
\and G.~Leroy\orcid{0009-0004-2523-4425}\inst{\ref{aff131},\ref{aff69}}
\and G.~F.~Lesci\orcid{0000-0002-4607-2830}\inst{\ref{aff68},\ref{aff14}}
\and J.~Lesgourgues\orcid{0000-0001-7627-353X}\inst{\ref{aff144}}
\and T.~I.~Liaudat\orcid{0000-0002-9104-314X}\inst{\ref{aff145}}
\and L.~Linke\orcid{0000-0002-2622-8113}\inst{\ref{aff88}}
\and S.~J.~Liu\orcid{0000-0001-7680-2139}\inst{\ref{aff43}}
\and F.~Mannucci\orcid{0000-0002-4803-2381}\inst{\ref{aff146}}
\and F.~R.~Marleau\orcid{0000-0002-1442-2947}\inst{\ref{aff88}}
\and C.~J.~A.~P.~Martins\orcid{0000-0002-4886-9261}\inst{\ref{aff147},\ref{aff148}}
\and M.~Miluzio\inst{\ref{aff12},\ref{aff149}}
\and G.~Morgante\inst{\ref{aff14}}
\and S.~Nadathur\orcid{0000-0001-9070-3102}\inst{\ref{aff120}}
\and K.~Naidoo\orcid{0000-0002-9182-1802}\inst{\ref{aff120},\ref{aff57}}
\and A.~Navarro-Alsina\orcid{0000-0002-3173-2592}\inst{\ref{aff66}}
\and S.~Nesseris\orcid{0000-0002-0567-0324}\inst{\ref{aff84}}
\and L.~Nicastro\orcid{0000-0001-8534-6788}\inst{\ref{aff14}}
\and F.~Oppizzi\orcid{0000-0003-3904-8370}\inst{\ref{aff2},\ref{aff41},\ref{aff83}}
\and F.~Pace\orcid{0000-0001-8039-0480}\inst{\ref{aff21},\ref{aff22},\ref{aff23},\ref{aff150}}
\and D.~Paoletti\orcid{0000-0003-4761-6147}\inst{\ref{aff14},\ref{aff44}}
\and G.~Parimbelli\orcid{0000-0002-2539-2472}\inst{\ref{aff28},\ref{aff15}}
\and F.~Passalacqua\orcid{0000-0002-8606-4093}\inst{\ref{aff41}}
\and K.~Paterson\orcid{0000-0001-8340-3486}\inst{\ref{aff57}}
\and L.~Patrizii\inst{\ref{aff17}}
\and C.~Pattison\orcid{0000-0003-3272-2617}\inst{\ref{aff120}}
\and R.~Paviot\orcid{0009-0002-8108-3460}\inst{\ref{aff79},\ref{aff81}}
\and A.~Pisani\orcid{0000-0002-6146-4437}\inst{\ref{aff42}}
\and D.~Potter\orcid{0000-0002-0757-5195}\inst{\ref{aff151}}
\and G.~W.~Pratt\inst{\ref{aff79}}
\and S.~Quai\orcid{0000-0002-0449-8163}\inst{\ref{aff68},\ref{aff14}}
\and M.~Radovich\orcid{0000-0002-3585-866X}\inst{\ref{aff18}}
\and G.~Rodighiero\orcid{0000-0002-9415-2296}\inst{\ref{aff83},\ref{aff18}}
\and W.~Roster\orcid{0000-0002-9149-6528}\inst{\ref{aff47}}
\and S.~Sacquegna\orcid{0000-0002-8433-6630}\inst{\ref{aff152}}
\and M.~Sahl\'en\orcid{0000-0003-0973-4804}\inst{\ref{aff153}}
\and D.~B.~Sanders\orcid{0000-0002-1233-9998}\inst{\ref{aff123}}
\and A.~Schneider\orcid{0000-0001-7055-8104}\inst{\ref{aff151}}
\and D.~Sciotti\orcid{0009-0008-4519-2620}\inst{\ref{aff30},\ref{aff67}}
\and E.~Sellentin\orcid{0009-0002-2655-3458}\inst{\ref{aff154},\ref{aff24}}
\and S.~Serjeant\orcid{0000-0002-0517-7943}\inst{\ref{aff155}}
\and L.~C.~Smith\orcid{0000-0002-3259-2771}\inst{\ref{aff156}}
\and J.~G.~Sorce\orcid{0000-0002-2307-2432}\inst{\ref{aff157},\ref{aff11}}
\and I.~Szapudi\orcid{0000-0003-2274-0301}\inst{\ref{aff123}}
\and M.~Talia\orcid{0000-0003-4352-2063}\inst{\ref{aff68},\ref{aff14}}
\and K.~Tanidis\orcid{0000-0001-9843-5130}\inst{\ref{aff158}}
\and C.~Tao\orcid{0000-0001-7961-8177}\inst{\ref{aff42}}
\and F.~Tarsitano\orcid{0000-0002-5919-0238}\inst{\ref{aff159},\ref{aff160},\ref{aff39}}
\and G.~Testera\orcid{0000-0003-2970-766X}\inst{\ref{aff2}}
\and R.~Teyssier\orcid{0000-0001-7689-0933}\inst{\ref{aff161}}
\and S.~Tosi\orcid{0000-0002-7275-9193}\inst{\ref{aff1},\ref{aff3},\ref{aff2}}
\and A.~Troja\orcid{0000-0003-0239-4595}\inst{\ref{aff6}}
\and C.~Uhlemann\orcid{0000-0001-7831-1579}\inst{\ref{aff162},\ref{aff75}}
\and C.~Valieri\inst{\ref{aff17}}
\and A.~Venhola\orcid{0000-0001-6071-4564}\inst{\ref{aff163}}
\and D.~Vergani\orcid{0000-0003-0898-2216}\inst{\ref{aff14}}
\and G.~Verza\orcid{0000-0002-1886-8348}\thanks{Deceased}\inst{\ref{aff164},\ref{aff165}}
\and S.~Vinciguerra\orcid{0009-0005-4018-3184}\inst{\ref{aff40}}
\and M.~von~Wietersheim-Kramsta\orcid{0000-0003-4986-5091}\inst{\ref{aff69},\ref{aff131}}
\and N.~A.~Walton\orcid{0000-0003-3983-8778}\inst{\ref{aff156}}
\and L.~Wang\orcid{0000-0002-6736-9158}\inst{\ref{aff166},\ref{aff122}}
\and A.~H.~Wright\orcid{0000-0001-7363-7932}\inst{\ref{aff129}}}
										   
%%%% please do not edit the affiliation list -- contact ECEB Bureau for changes
\institute{Dipartimento di Fisica, Universit\`a di Genova, Via Dodecaneso 33, 16146, Genova, Italy\label{aff1}
\and
INFN-Sezione di Genova, Via Dodecaneso 33, 16146, Genova, Italy\label{aff2}
\and
INAF-Osservatorio Astronomico di Brera, Via Brera 28, 20122 Milano, Italy\label{aff3}
\and
Universit\'e Claude Bernard Lyon 1, CNRS/IN2P3, IP2I Lyon, UMR 5822, Villeurbanne, F-69100, France\label{aff4}
\and
Dipartimento di Fisica - Sezione di Astronomia, Universit\`a di Trieste, Via Tiepolo 11, 34131 Trieste, Italy\label{aff5}
\and
INAF-Osservatorio Astronomico di Trieste, Via G. B. Tiepolo 11, 34143 Trieste, Italy\label{aff6}
\and
INFN, Sezione di Trieste, Via Valerio 2, 34127 Trieste TS, Italy\label{aff7}
\and
IFPU, Institute for Fundamental Physics of the Universe, via Beirut 2, 34151 Trieste, Italy\label{aff8}
\and
Univ. Grenoble Alpes, CNRS, Grenoble INP, LPSC-IN2P3, 53, Avenue des Martyrs, 38000, Grenoble, France\label{aff9}
\and
Institute for Astronomy, University of Edinburgh, Royal Observatory, Blackford Hill, Edinburgh EH9 3HJ, UK\label{aff10}
\and
Universit\'e Paris-Saclay, CNRS, Institut d'astrophysique spatiale, 91405, Orsay, France\label{aff11}
\and
ESAC/ESA, Camino Bajo del Castillo, s/n., Urb. Villafranca del Castillo, 28692 Villanueva de la Ca\~nada, Madrid, Spain\label{aff12}
\and
School of Mathematics and Physics, University of Surrey, Guildford, Surrey, GU2 7XH, UK\label{aff13}
\and
INAF-Osservatorio di Astrofisica e Scienza dello Spazio di Bologna, Via Piero Gobetti 93/3, 40129 Bologna, Italy\label{aff14}
\and
SISSA, International School for Advanced Studies, Via Bonomea 265, 34136 Trieste TS, Italy\label{aff15}
\and
Dipartimento di Fisica e Astronomia, Universit\`a di Bologna, Via Gobetti 93/2, 40129 Bologna, Italy\label{aff16}
\and
INFN-Sezione di Bologna, Viale Berti Pichat 6/2, 40127 Bologna, Italy\label{aff17}
\and
INAF-Osservatorio Astronomico di Padova, Via dell'Osservatorio 5, 35122 Padova, Italy\label{aff18}
\and
Department of Physics "E. Pancini", University Federico II, Via Cinthia 6, 80126, Napoli, Italy\label{aff19}
\and
INAF-Osservatorio Astronomico di Capodimonte, Via Moiariello 16, 80131 Napoli, Italy\label{aff20}
\and
Dipartimento di Fisica, Universit\`a degli Studi di Torino, Via P. Giuria 1, 10125 Torino, Italy\label{aff21}
\and
INFN-Sezione di Torino, Via P. Giuria 1, 10125 Torino, Italy\label{aff22}
\and
INAF-Osservatorio Astrofisico di Torino, Via Osservatorio 20, 10025 Pino Torinese (TO), Italy\label{aff23}
\and
Leiden Observatory, Leiden University, Einsteinweg 55, 2333 CC Leiden, The Netherlands\label{aff24}
\and
INAF-IASF Milano, Via Alfonso Corti 12, 20133 Milano, Italy\label{aff25}
\and
Centro de Investigaciones Energ\'eticas, Medioambientales y Tecnol\'ogicas (CIEMAT), Avenida Complutense 40, 28040 Madrid, Spain\label{aff26}
\and
Port d'Informaci\'{o} Cient\'{i}fica, Campus UAB, C. Albareda s/n, 08193 Bellaterra (Barcelona), Spain\label{aff27}
\and
Institute of Space Sciences (ICE, CSIC), Campus UAB, Carrer de Can Magrans, s/n, 08193 Barcelona, Spain\label{aff28}
\and
Institut d'Estudis Espacials de Catalunya (IEEC),  Edifici RDIT, Campus UPC, 08860 Castelldefels, Barcelona, Spain\label{aff29}
\and
INAF-Osservatorio Astronomico di Roma, Via Frascati 33, 00078 Monteporzio Catone, Italy\label{aff30}
\and
INFN -- Sezione di Napoli, Via Cinthia 6, 80126, Napoli, Italy\label{aff31}
\and
Dipartimento di Fisica e Astronomia "Augusto Righi" - Alma Mater Studiorum Universit\`a di Bologna, Viale Berti Pichat 6/2, 40127 Bologna, Italy\label{aff32}
\and
Instituto de Astrof\'{\i}sica de Canarias, E-38205 La Laguna, Tenerife, Spain\label{aff33}
\and
European Space Agency/ESRIN, Largo Galileo Galilei 1, 00044 Frascati, Roma, Italy\label{aff34}
\and
Institut de Ci\`{e}ncies del Cosmos (ICCUB), Universitat de Barcelona (IEEC-UB), Mart\'{i} i Franqu\`{e}s 1, 08028 Barcelona, Spain\label{aff35}
\and
Instituci\'o Catalana de Recerca i Estudis Avan\c{c}ats (ICREA), Passeig de Llu\'{\i}s Companys 23, 08010 Barcelona, Spain\label{aff36}
\and
Institut de Ciencies de l'Espai (IEEC-CSIC), Campus UAB, Carrer de Can Magrans, s/n Cerdanyola del Vall\'es, 08193 Barcelona, Spain\label{aff37}
\and
UCB Lyon 1, CNRS/IN2P3, IUF, IP2I Lyon, 4 rue Enrico Fermi, 69622 Villeurbanne, France\label{aff38}
\and
Department of Astronomy, University of Geneva, ch. d'Ecogia 16, 1290 Versoix, Switzerland\label{aff39}
\and
Aix-Marseille Universit\'e, CNRS, CNES, LAM, Marseille, France\label{aff40}
\and
INFN-Padova, Via Marzolo 8, 35131 Padova, Italy\label{aff41}
\and
Aix-Marseille Universit\'e, CNRS/IN2P3, CPPM, Marseille, France\label{aff42}
\and
INAF-Istituto di Astrofisica e Planetologia Spaziali, via del Fosso del Cavaliere, 100, 00100 Roma, Italy\label{aff43}
\and
INFN-Bologna, Via Irnerio 46, 40126 Bologna, Italy\label{aff44}
\and
Department of Physics, University of Oxford, Keble Road, Oxford OX1 3RH, UK\label{aff45}
\and
University Observatory, LMU Faculty of Physics, Scheinerstr.~1, 81679 Munich, Germany\label{aff46}
\and
Max Planck Institute for Extraterrestrial Physics, Giessenbachstr. 1, 85748 Garching, Germany\label{aff47}
\and
Universit\"ats-Sternwarte M\"unchen, Fakult\"at f\"ur Physik, Ludwig-Maximilians-Universit\"at M\"unchen, Scheinerstr.~1, 81679 M\"unchen, Germany\label{aff48}
\and
Dipartimento di Fisica "Aldo Pontremoli", Universit\`a degli Studi di Milano, Via Celoria 16, 20133 Milano, Italy\label{aff49}
\and
INFN-Sezione di Milano, Via Celoria 16, 20133 Milano, Italy\label{aff50}
\and
Institute of Theoretical Astrophysics, University of Oslo, P.O. Box 1029 Blindern, 0315 Oslo, Norway\label{aff51}
\and
Caltech/IPAC, 1200 E. California Blvd., Pasadena, CA 91125, USA\label{aff52}
\and
Jet Propulsion Laboratory, California Institute of Technology, 4800 Oak Grove Drive, Pasadena, CA, 91109, USA\label{aff53}
\and
Felix Hormuth Engineering, Goethestr. 17, 69181 Leimen, Germany\label{aff54}
\and
Technical University of Denmark, Elektrovej 327, 2800 Kgs. Lyngby, Denmark\label{aff55}
\and
Cosmic Dawn Center (DAWN), Denmark\label{aff56}
\and
Max-Planck-Institut f\"ur Astronomie, K\"onigstuhl 17, 69117 Heidelberg, Germany\label{aff57}
\and
NASA Goddard Space Flight Center, Greenbelt, MD 20771, USA\label{aff58}
\and
Department of Physics and Astronomy, University College London, Gower Street, London WC1E 6BT, UK\label{aff59}
\and
Universit\'e de Gen\`eve, D\'epartement de Physique Th\'eorique and Centre for Astroparticle Physics, 24 quai Ernest-Ansermet, CH-1211 Gen\`eve 4, Switzerland\label{aff60}
\and
Department of Physics, P.O. Box 64, University of Helsinki, 00014 Helsinki, Finland\label{aff61}
\and
Helsinki Institute of Physics, Gustaf H{\"a}llstr{\"o}min katu 2, University of Helsinki, 00014 Helsinki, Finland\label{aff62}
\and
Laboratoire d'etude de l'Univers et des phenomenes eXtremes, Observatoire de Paris, Universit\'e PSL, Sorbonne Universit\'e, CNRS, 92190 Meudon, France\label{aff63}
\and
SKAO, Jodrell Bank, Lower Withington, Macclesfield SK11 9FT, UK\label{aff64}
\and
Centre de Calcul de l'IN2P3/CNRS, 21 avenue Pierre de Coubertin 69627 Villeurbanne Cedex, France\label{aff65}
\and
Universit\"at Bonn, Argelander-Institut f\"ur Astronomie, Auf dem H\"ugel 71, 53121 Bonn, Germany\label{aff66}
\and
INFN-Sezione di Roma, Piazzale Aldo Moro, 2 - c/o Dipartimento di Fisica, Edificio G. Marconi, 00185 Roma, Italy\label{aff67}
\and
Dipartimento di Fisica e Astronomia "Augusto Righi" - Alma Mater Studiorum Universit\`a di Bologna, via Piero Gobetti 93/2, 40129 Bologna, Italy\label{aff68}
\and
Department of Physics, Institute for Computational Cosmology, Durham University, South Road, Durham, DH1 3LE, UK\label{aff69}
\and
University of Applied Sciences and Arts of Northwestern Switzerland, School of Engineering, 5210 Windisch, Switzerland\label{aff70}
\and
Institute of Physics, Laboratory of Astrophysics, Ecole Polytechnique F\'ed\'erale de Lausanne (EPFL), Observatoire de Sauverny, 1290 Versoix, Switzerland\label{aff71}
\and
Telespazio UK S.L. for European Space Agency (ESA), Camino bajo del Castillo, s/n, Urbanizacion Villafranca del Castillo, Villanueva de la Ca\~nada, 28692 Madrid, Spain\label{aff72}
\and
Institut de F\'{i}sica d'Altes Energies (IFAE), The Barcelona Institute of Science and Technology, Campus UAB, 08193 Bellaterra (Barcelona), Spain\label{aff73}
\and
European Space Agency/ESTEC, Keplerlaan 1, 2201 AZ Noordwijk, The Netherlands\label{aff74}
\and
School of Mathematics, Statistics and Physics, Newcastle University, Herschel Building, Newcastle-upon-Tyne, NE1 7RU, UK\label{aff75}
\and
Waterloo Centre for Astrophysics, University of Waterloo, Waterloo, Ontario N2L 3G1, Canada\label{aff76}
\and
Department of Physics and Astronomy, University of Waterloo, Waterloo, Ontario N2L 3G1, Canada\label{aff77}
\and
Perimeter Institute for Theoretical Physics, Waterloo, Ontario N2L 2Y5, Canada\label{aff78}
\and
Universit\'e Paris-Saclay, Universit\'e Paris Cit\'e, CEA, CNRS, AIM, 91191, Gif-sur-Yvette, France\label{aff79}
\and
Space Science Data Center, Italian Space Agency, via del Politecnico snc, 00133 Roma, Italy\label{aff80}
\and
Centre National d'Etudes Spatiales -- Centre spatial de Toulouse, 18 avenue Edouard Belin, 31401 Toulouse Cedex 9, France\label{aff81}
\and
Institute of Space Science, Str. Atomistilor, nr. 409 M\u{a}gurele, Ilfov, 077125, Romania\label{aff82}
\and
Dipartimento di Fisica e Astronomia "G. Galilei", Universit\`a di Padova, Via Marzolo 8, 35131 Padova, Italy\label{aff83}
\and
Instituto de F\'isica Te\'orica UAM-CSIC, Campus de Cantoblanco, 28049 Madrid, Spain\label{aff84}
\and
Institut de Recherche en Astrophysique et Plan\'etologie (IRAP), Universit\'e de Toulouse, CNRS, UPS, CNES, 14 Av. Edouard Belin, 31400 Toulouse, France\label{aff85}
\and
Universit\'e St Joseph; Faculty of Sciences, Beirut, Lebanon\label{aff86}
\and
Departamento de F\'isica, FCFM, Universidad de Chile, Blanco Encalada 2008, Santiago, Chile\label{aff87}
\and
Universit\"at Innsbruck, Institut f\"ur Astro- und Teilchenphysik, Technikerstr. 25/8, 6020 Innsbruck, Austria\label{aff88}
\and
Department of Physics and Helsinki Institute of Physics, Gustaf H\"allstr\"omin katu 2, University of Helsinki, 00014 Helsinki, Finland\label{aff89}
\and
Departamento de F\'isica, Faculdade de Ci\^encias, Universidade de Lisboa, Edif\'icio C8, Campo Grande, PT1749-016 Lisboa, Portugal\label{aff90}
\and
Instituto de Astrof\'isica e Ci\^encias do Espa\c{c}o, Faculdade de Ci\^encias, Universidade de Lisboa, Tapada da Ajuda, 1349-018 Lisboa, Portugal\label{aff91}
\and
Mullard Space Science Laboratory, University College London, Holmbury St Mary, Dorking, Surrey RH5 6NT, UK\label{aff92}
\and
Cosmic Dawn Center (DAWN)\label{aff93}
\and
Niels Bohr Institute, University of Copenhagen, Jagtvej 128, 2200 Copenhagen, Denmark\label{aff94}
\and
Universidad Polit\'ecnica de Cartagena, Departamento de Electr\'onica y Tecnolog\'ia de Computadoras,  Plaza del Hospital 1, 30202 Cartagena, Spain\label{aff95}
\and
European University of Technology EUt+, European Union\label{aff96}
\and
Dipartimento di Fisica e Scienze della Terra, Universit\`a degli Studi di Ferrara, Via Giuseppe Saragat 1, 44122 Ferrara, Italy\label{aff97}
\and
Istituto Nazionale di Fisica Nucleare, Sezione di Ferrara, Via Giuseppe Saragat 1, 44122 Ferrara, Italy\label{aff98}
\and
INAF, Istituto di Radioastronomia, Via Piero Gobetti 101, 40129 Bologna, Italy\label{aff99}
\and
Astronomical Observatory of the Autonomous Region of the Aosta Valley (OAVdA), Loc. Lignan 39, I-11020, Nus (Aosta Valley), Italy\label{aff100}
\and
Universit\'e C\^{o}te d'Azur, Observatoire de la C\^{o}te d'Azur, CNRS, Laboratoire Lagrange, Bd de l'Observatoire, CS 34229, 06304 Nice cedex 4, France\label{aff101}
\and
Department of Mathematical Physics, Institute of Physics, University of S\~ao Paulo, R. do Mat\~ao 1371, 05508-090, S\~ao Paulo, SP, Brazil\label{aff102}
\and
Dipartimento di Fisica, Sapienza Universit\`a di Roma, Piazzale Aldo Moro 2, 00185 Roma, Italy\label{aff103}
\and
Department of Mathematics and Physics E. De Giorgi, University of Salento, Via per Arnesano, CP-I93, 73100, Lecce, Italy\label{aff104}
\and
INFN, Sezione di Lecce, Via per Arnesano, CP-193, 73100, Lecce, Italy\label{aff105}
\and
INAF-Sezione di Lecce, c/o Dipartimento Matematica e Fisica, Via per Arnesano, 73100, Lecce, Italy\label{aff106}
\and
Institut d'Astrophysique de Paris, 98bis Boulevard Arago, 75014, Paris, France\label{aff107}
\and
ICL, Junia, Universit\'e Catholique de Lille, LITL, 59000 Lille, France\label{aff108}
\and
ICSC - Centro Nazionale di Ricerca in High Performance Computing, Big Data e Quantum Computing, Via Magnanelli 2, Bologna, Italy\label{aff109}
\and
CERCA/ISO, Department of Physics, Case Western Reserve University, 10900 Euclid Avenue, Cleveland, OH 44106, USA\label{aff110}
\and
Laboratoire Univers et Th\'eorie, Observatoire de Paris, Universit\'e PSL, Universit\'e Paris Cit\'e, CNRS, 92190 Meudon, France\label{aff111}
\and
Departamento de F{\'\i}sica Fundamental. Universidad de Salamanca. Plaza de la Merced s/n. 37008 Salamanca, Spain\label{aff112}
\and
Universiteit Gent, Department of Physics and Astronomy, Proeftuinstraat 86 N3, 9000 Ghent, Belgium
\label{aff113}
\and
Aix-Marseille Universit\'e, Universit\'e de Toulon, CNRS, CPT, Marseille, France\label{aff114}
\and
Universit\'e de Strasbourg, CNRS, Observatoire astronomique de Strasbourg, UMR 7550, 67000 Strasbourg, France\label{aff115}
\and
Center for Data-Driven Discovery, Kavli IPMU (WPI), UTIAS, The University of Tokyo, Kashiwa, Chiba 277-8583, Japan\label{aff116}
\and
Jodrell Bank Centre for Astrophysics, Department of Physics and Astronomy, University of Manchester, Oxford Road, Manchester M13 9PL, UK\label{aff117}
\and
California Institute of Technology, 1200 E California Blvd, Pasadena, CA 91125, USA\label{aff118}
\and
Department of Physics, Astronomy and Mathematics, University of Hertfordshire, College Lane, Hatfield AL10 9AB, UK\label{aff119}
\and
Institute of Cosmology and Gravitation, University of Portsmouth, Portsmouth PO1 3FX, UK\label{aff120}
\and
Department of Physics \& Astronomy, University of California Irvine, Irvine CA 92697, USA\label{aff121}
\and
Kapteyn Astronomical Institute, University of Groningen, PO Box 800, 9700 AV Groningen, The Netherlands\label{aff122}
\and
Institute for Astronomy, University of Hawaii, 2680 Woodlawn Drive, Honolulu, HI 96822, USA\label{aff123}
\and
Departamento F\'isica Aplicada, Universidad Polit\'ecnica de Cartagena, Campus Muralla del Mar, 30202 Cartagena, Murcia, Spain\label{aff124}
\and
Instituto de F\'isica de Cantabria, Edificio Juan Jord\'a, Avenida de los Castros, 39005 Santander, Spain\label{aff125}
\and
Universit\'e Paris Cit\'e, CNRS, Astroparticule et Cosmologie, 75013 Paris, France\label{aff126}
\and
Department of Computer Science, Aalto University, PO Box 15400, Espoo, FI-00 076, Finland\label{aff127}
\and
Universidad de La Laguna, Dpto. Astrof\'\i sica, E-38206 La Laguna, Tenerife, Spain\label{aff128}
\and
Ruhr University Bochum, Faculty of Physics and Astronomy, Astronomical Institute (AIRUB), German Centre for Cosmological Lensing (GCCL), 44780 Bochum, Germany\label{aff129}
\and
DARK, Niels Bohr Institute, University of Copenhagen, Jagtvej 155, 2200 Copenhagen, Denmark\label{aff130}
\and
Department of Physics, Centre for Extragalactic Astronomy, Durham University, South Road, Durham, DH1 3LE, UK\label{aff131}
\and
Astrophysics Research Centre, University of KwaZulu-Natal, Westville Campus, Durban 4041, South Africa\label{aff132}
\and
School of Mathematics, Statistics \& Computer Science, University of KwaZulu-Natal, Westville Campus, Durban 4041, South Africa\label{aff133}
\and
Department of Physics and Astronomy, Vesilinnantie 5, University of Turku, 20014 Turku, Finland\label{aff134}
\and
Finnish Centre for Astronomy with ESO (FINCA), Quantum, Vesilinnantie 5, University of Turku, 20014 Turku, Finland\label{aff135}
\and
Serco for European Space Agency (ESA), Camino bajo del Castillo, s/n, Urbanizacion Villafranca del Castillo, Villanueva de la Ca\~nada, 28692 Madrid, Spain\label{aff136}
\and
ARC Centre of Excellence for Dark Matter Particle Physics, Melbourne, Australia\label{aff137}
\and
Centre for Astrophysics \& Supercomputing, Swinburne University of Technology,  Hawthorn, Victoria 3122, Australia\label{aff138}
\and
Department of Physics and Astronomy, University of the Western Cape, Bellville, Cape Town, 7535, South Africa\label{aff139}
\and
MTA-CSFK Lend\"ulet Large-Scale Structure Research Group, Konkoly-Thege Mikl\'os \'ut 15-17, H-1121 Budapest, Hungary\label{aff140}
\and
Konkoly Observatory, HUN-REN CSFK, MTA Centre of Excellence, Budapest, Konkoly Thege Mikl\'os {\'u}t 15-17. H-1121, Hungary\label{aff141}
\and
Institut d'Astrophysique de Paris, UMR 7095, CNRS, and Sorbonne Universit\'e, 98 bis boulevard Arago, 75014 Paris, France\label{aff142}
\and
Brazilian Center for Research in Physics (CBPF), Dr. Xavier Sigaud st. 150, zip 22290-180, Rio de Janeiro, RJ, Brazil\label{aff143}
\and
Institute for Theoretical Particle Physics and Cosmology (TTK), RWTH Aachen University, 52056 Aachen, Germany\label{aff144}
\and
IRFU, CEA, Universit\'e Paris-Saclay 91191 Gif-sur-Yvette Cedex, France\label{aff145}
\and
INAF-Osservatorio Astrofisico di Arcetri, Largo E. Fermi 5, 50125, Firenze, Italy\label{aff146}
\and
Centro de Astrof\'{\i}sica da Universidade do Porto, Rua das Estrelas, 4150-762 Porto, Portugal\label{aff147}
\and
Instituto de Astrof\'isica e Ci\^encias do Espa\c{c}o, Universidade do Porto, CAUP, Rua das Estrelas, PT4150-762 Porto, Portugal\label{aff148}
\and
HE Space for European Space Agency (ESA), Camino bajo del Castillo, s/n, Urbanizacion Villafranca del Castillo, Villanueva de la Ca\~nada, 28692 Madrid, Spain\label{aff149}
\and
Instituto de Astrof\'isica e Ci\^encias do Espa\c{c}o, Faculdade de Ci\^encias, Universidade de Lisboa, Campo Grande, 1749-016 Lisboa, Portugal\label{aff150}
\and
Department of Astrophysics, University of Zurich, Winterthurerstrasse 190, 8057 Zurich, Switzerland\label{aff151}
\and
INAF - Osservatorio Astronomico d'Abruzzo, Via Maggini, 64100, Teramo, Italy\label{aff152}
\and
Theoretical astrophysics, Department of Physics and Astronomy, Uppsala University, Box 516, 751 37 Uppsala, Sweden\label{aff153}
\and
Mathematical Institute, University of Leiden, Einsteinweg 55, 2333 CA Leiden, The Netherlands\label{aff154}
\and
School of Physical Sciences, The Open University, Milton Keynes, MK7 6AA, UK\label{aff155}
\and
Institute of Astronomy, University of Cambridge, Madingley Road, Cambridge CB3 0HA, UK\label{aff156}
\and
Univ. Lille, CNRS, Centrale Lille, UMR 9189 CRIStAL, 59000 Lille, France\label{aff157}
\and
Center for Astrophysics and Cosmology, University of Nova Gorica, Nova Gorica, Slovenia\label{aff158}
\and
Kobayashi-Maskawa Institute for the Origin of Particles and the Universe, Nagoya University, Chikusa-ku, Nagoya, 464-8602, Japan\label{aff159}
\and
Institute for Particle Physics and Astrophysics, Dept. of Physics, ETH Zurich, Wolfgang-Pauli-Strasse 27, 8093 Zurich, Switzerland\label{aff160}
\and
Department of Astrophysical Sciences, Peyton Hall, Princeton University, Princeton, NJ 08544, USA\label{aff161}
\and
Fakult\"at f\"ur Physik, Universit\"at Bielefeld, Postfach 100131, 33501 Bielefeld, Germany\label{aff162}
\and
Space physics and astronomy research unit, University of Oulu, Pentti Kaiteran katu 1, FI-90014 Oulu, Finland\label{aff163}
\and
International Centre for Theoretical Physics (ICTP), Strada Costiera 11, 34151 Trieste, Italy\label{aff164}
\and
Center for Computational Astrophysics, Flatiron Institute, 162 5th Avenue, 10010, New York, NY, USA\label{aff165}
\and
SRON Netherlands Institute for Space Research, Landleven 12, 9747 AD, Groningen, The Netherlands\label{aff166}}      

% 
% Put your abstract here:
%
\abstract
{
The \Euclid spectroscopic survey will measure galaxy clustering with unprecedented precision, requiring a stringent control of observational and instrumental systematics. Among these, star-related effects may contaminate the spectroscopic images through photometric persistence and through imperfect masking of the star images.
%}
%{ 
We characterized the imprint of star-related effects in the spectroscopic exposures and quantified their impact on the galaxy clustering analyses. 
If these effects produce characteristic angular features, they can be identified and characterized
in the data, thereby enabling an assessment of their impact on cosmological measurements.
%}
%{

We used angular and spatial auto and cross-correlation statistics to quantify these effects. For the galaxies we used
mock spectroscopic catalogues extracted from the EuclidLargeMocks. We focused on a $330 \sqdeg$ region of the Euclid Wide Survey (EWS).
For the stars we considered the real  \Gaia\ and \twomass\ star catalogues in the same area. To simulate photometric persistence, we implemented a simplified detector-level model calibrated on the spectroscopic measurements and we 
modulated the strength of the effect to introduce varying fractions of `interlopers' into the mock galaxy catalogues.
For the stellar mask, we modelled potential inconsistencies between the mask used to remove regions around stars in the data and the corresponding mask applied to the random catalogue used to quantify the survey selection function. 
We measured the star--star, galaxy--galaxy, and star--galaxy angular correlation functions using the Landy--Szalay (LS) estimator, and quantified deviations from the expected null star–galaxy correlation with a dedicated metric. 
Finally, we evaluated the impact of these systematics on large scales through the three-dimensional two-point correlation function (2PCF).
%}
%{

Our results show that photometric persistence produces a characteristic feature in the star--galaxy angular cross-correlation at scales of order $100 \arcsec$, corresponding to the scale of the \Euclid dithering pattern and grism dispersion geometry. It also induces a spurious, positive cross-correlation signal that remains approximately constant in amplitude up to $1^\circ$. We found that star–galaxy cross-correlation analysis is a sensitive tool for detecting residual persistence, capable of revealing spurious contamination levels as low as 10\% in a Data Release 1 spectroscopic catalogue. In contrast, mismatches in the stellar masking produce strong small-scale angular signatures but have a negligible impact on the large-scale 2PCF under realistic EWS conditions.}
%{}

%
% Provide up to five key words:
%
%    from the list in
%     https://www.aanda.org/for-authors/latex-issues/information-files#pop}
%
% Add short versions of title and author list for page headings
%
\titlerunning{A two-point correlation approach to diagnosing star-related systematics} 
%in the \Euclid spectroscopic survey}
\authorrunning{Euclid Collaboration: I.~Risso et al.}

\maketitle
%
%-------------------------------------------------------------------
%
%
%   Start the main text of your paper here
%

\section{\label{sc:Intro}Introduction}

The \Euclid mission \citep{EuclidSkyOverview} is designed to investigate the nature of dark energy and dark matter through high-precision measurements of galaxy clustering and weak gravitational lensing. For galaxy clustering, the three-dimensional galaxy two-point correlation function (2PCF) is the key statistic used to infer the cosmology underlying the observed galaxy distribution. This measurement relies on the spectroscopic sample of \ha\ emitting galaxies over a wide redshift range, obtained with the Near-Infrared Spectrometer and Photometer instrument (NISP, \citealp{EuclidSkyNISP}).
NISP performs slitless spectroscopy in the near-infrared using two red grisms, which are observed with different dispersion orientations, combined with a sequence of photometric exposures in three broad near-infrared bands \YE (950--1212\,\si{nm}), \JE (1168--1567\,\si{nm}), and \HE (1522--2021\,\si{nm}). In the Euclid Wide Survey (EWS, \citealp{Scaramella-EP1}), the spectroscopic observations cover the wavelength range 1206--1892\,\si{nm}, enabling the detection of \ha\ emission lines across the redshift range $0.84 \leq z \leq 1.88$ \citep{EuclidSkyOverview}. The combination of multiple grism orientations and dithered observations, following the reference observation sequence adopted in the EWS, is designed to mitigate spectral overlaps and improve redshift measurements.

Over the past two decades, the control of observational systematics has become an increasingly important aspect of large-scale structure analyses. Variations in stellar density, seeing, extinction, and survey depth imprint themselves on the measured galaxy density field leading to  biases in clustering measurements \citep{Ross_2011, Ross_2012, Ho_2012, Guzzo_2014, Crocee_2016}. Such spatially varying selection effects have become a limiting source of systematic uncertainty, motivating the development of dedicated diagnostic and mitigation strategies \citep{Johnston_2021, RosadoMarin_2025, Krolewski_2025}. Moreover, spectroscopic surveys are affected by targetting biases \citep{Bianchi_2017, Pezzotta_2017} and redshift measurement failures \citep{Yu_2025}.
 
Likewise, the scientific return of \Euclid\ critically depends on the accurate characterization and control of observational and instrumental systematic effects \citep{EP-Monaco2}, as these can bias clustering measurements, distort the baryon acoustic oscillation (BAO) feature, and introduce artificial scale-dependent signatures.

%In this work, 
Unlike the aforementioned works, which primarily focus on modelling or correcting systematics, here we investigate the use of angular two-point statistics themselves as a diagnostic tool to identify residual star-related systematics that may affect the \Euclid spectroscopic catalogues during the first public data release (DR1).
Specifically, we focus on two selection effects that relate to stars: contamination caused by the persistence of sources in the photometric images to the following spectroscopic exposures and true targets that are lost near to bright stars.
%mismatch between the \textcolor{magenta}{effective} stellar mask
%\sout{applied to} 
%in the spectroscopic data and random catalogues.

Photometric persistence occurs when charge carriers generated during a photometric exposure are not fully cleared before the subsequent spectroscopic exposure \citep{persistence-theory}. The residual signal may be misinterpreted as an emission feature if it happens to overlap with a galaxy spectrum, potentially leading to catastrophic redshift assignments and to the inclusion of spurious sources in the spectroscopic catalogue (the `interlopers', whose impact on \Euclid clustering analysis is described in \citealt{EP-Risso}). Since persistence signals are not randomly distributed but follow the \Euclid observing sequence and dithering strategy in the EWS, they are expected to leave characteristic angular imprints in the angular two-point statistics. A comprehensive overview of persistence modelling for DR1 is presented in Euclid Collaboration: Kubik et al. (in prep.).
%\citet{DR1-TP038}.
%\textcolor{magenta}{In this work, we do not attempt to estimate the level of residual persistence expected in the DR1 catalogue. Rather, we investigate how residual persistence, if present after the mitigation procedures, would manifest itself in the angular clustering statistics.}
For DR1, the NIR \citep[Near InfraRed imaging data,][]{Q1-TP003} Processing Function (PF) masked pixels in the spectroscopy predicted to be significantly affected by persistence from the photometric images 
%\citep{DR1-TP038}
Euclid Collaboration: Kubik et al. (in prep.). In this work, we investigate the impact of residual persistence signals that could remain after this mitigation, for example because of simplifying assumptions in the adopted temporal masking model or because low-level persistence is not explicitly masked.

The second systematic concerns the star masking procedure. In the \Euclid pipeline, polygon masks are defined around bright stars to reduce the rate of spurious detections originating from diffraction spikes by the MER \citep[MERged data,][]{Q1-TP004} PF. The random catalogue is constructed by the VMSP PF (Spectroscopic Visibility Mask pipeline, Euclid Collaboration: Granett et al. 2026, in prep.), which applies a pixelized version of the star mask. However, the detection efficiency of background galaxies also drops near to bright stars due to the high background and point spread function leaving empty regions in the galaxy catalogue. The size of these holes is generally larger than the explicitly masked regions, and the size of the mask must be calibrated using statistical techniques including the star-galaxy cross-correlation function.
% , while the masking in the random catalogue is implemented by the VMSP PF (Spectroscopic Visibility Mask pipeline, Euclid Collaboration: Granett et al. 2026, in prep.),
%\citep[Spectroscopic Visibility Mask pipeline,][Euclid Collaboration: Granett et al, in preparation]{}, 
% relying on a pixelized coverage map produced by the VMPZ PF \citep[Photo-$z$ Visibility Mask pipeline,][]{DR1-TP030} based on the MER masking. \textcolor{magenta}{This may introduce little geometrical mismatches in the masked area in the two catalogues. In addition, there may be a further collateral galaxy masking present in the data around stars but not in the random catalogue, not due to the different applied mask itself but to physical effects which cannot be traced in the random catalogue.}
Any residual mismatch between the effective masked area in the data and in the random catalogue introduces artificial correlations between stars and galaxies that must be assessed.

Angular two-point statistics provide a powerful and direct diagnostic tool for detecting such effects. In the absence of systematics, we do not expect any intrinsic angular cross-correlation between stars and spectroscopic galaxies. Therefore, a statistically significant deviation from zero in the star–galaxy angular cross-correlation function indicates the presence of residual contamination.

In this work, we used angular two-point statistics to detect the presence of star-related contamination through the star–galaxy cross-correlation, and we quantified its statistical significance in view of the \Euclid DR1 data.
We performed our analysis using the EuclidLargeMocks \citep[ELM,][]{EP-Monaco1} as reference spectroscopic galaxy catalogues and real star catalogues from \Gaia\ \citep{Gaia_2016A&A595A1G} and \twomass\ \citep{2mass_2006AJ....131.1163S} to compute cross-correlations and simulate star-related systematics.
%, restricting the study to a compact region in the South of the EWS DR1 footprint, available prior to the completion of the processing of the full DR1 field.
The mock galaxy catalogues were restricted to a compact region within the EWS footprint, named `S1 field', corresponding to an area that was extensively studied during the Euclid DR1 validation phase.
The use of mock catalogues allowed us to isolate and characterise the expected imprint of stellar contamination in a controlled setup. The application of these diagnostics to \Euclid observational data will be carried out in a dedicated DR1 paper.
%No \Euclid observational data were used in this analysis.

%\sout{For persistence, we constructed a simplified yet physically motivated model calibrated on in-flight detector-level measurements} \citep{EU-Kubik}, \sout{and we simulated contaminated catalogues with varying fractions of residual interlopers.  We expect the contamination fraction to vary with Galactic latitude, as it correlates with the stellar density, and with the density of spectra extracted through the spectroscopic pipeline. In DR1, persistence will be mitigated by the NIR PF} \citep[Near InfraRed imaging data,][]{Q1-TP003} \sout{by masking the pixels predicted to be significantly affected} \citep{DR1-TP038}. \sout{Here, we focus on the residual persistence signal that may remain because of incomplete or imperfect masking. For the star mask, we mimicked realistic mismatches between the \textcolor{magenta}{effective} angular size of the star masks in the data and random catalogues.}

Because of their regular angular pattern, these systematics may affect two-point clustering measurements out to large separations, potentially including the BAO peak. We therefore assessed their impact on the 2PCF.
The goal of this work is therefore twofold:
first of all, to provide robust diagnostics for detecting star-related systematics in early \Euclid data, and
second, to assess their potential impact on galaxy clustering analyses, establishing thresholds below which cosmological measurements remain unbiased.

The paper is structured as follows. In Sect.~\ref{sc:observation-strategy} we describe the dithered observation strategy adopted by \Euclid in the standard scientific observations, which is fundamental for the correct simulation of persistence. In Sect.~\ref{sc:data} we describe the mock galaxy catalogues and the real star catalogues we used. Section \ref{sc:tools} describes the clustering statistics adopted in this work, as well as the procedure we adopted to simulate spectra and to predict the angular location of the persistence signals. Section \ref{sc:persistence} is about persistence. After introducing its physical origin, we detail all the steps we followed in the simulation of star persistence and the approximations we made, concluding with the results quantifying the impact of persistence on the angular statistics. In Sect.~\ref{sc:starmask}, we assess the impact of the mismatch of the star mask in the spectroscopic data and random catalogues. In Sect.~\ref{sc:largescales}, we evaluate the impact of both systematic effects on the 2PCF, extending the analysis from the small angular scales where these systematics are identified to large spatial scales encompassing the BAO peak. We draw our conclusions in Sect.~\ref{sc:Conclusion}.

\section{\label{sc:observation-strategy}\Euclid survey strategy
}

We summarise here the aspects of the EWS that are relevant to this work. A clear understanding of how photometric and spectroscopic exposures alternate is essential to correctly interpret the results of this work and the procedure followed for the persistence analysis in particular.

A detailed description of the survey strategy can be found in \citet{Scaramella-EP1}, whose figure 8 shows the \Euclid reference observation sequence (ROS). We include the same figure here in
%Fig.~\ref{fig:surveysequence} of
Appendix \ref{app:ROS} for the reader's convenience. 
%\ref{fig:surveysequence} . 
Each observation of a sky area corresponding to the telescope field of view is made of four different pointings (or `dithers') interleaved by dither steps, characterized by small changes in the satellite's attitude. 
During each of the four dithers, both photometric and spectroscopic data are acquired through different kinds of exposures. In particular:
\begin{itemize}
    \item VIS photometric and near-infrared spectroscopic images are taken simultaneously. The spectroscopic exposure lasts 574 s;
    \item VIS closes its shutter. After rotating the filter and grism wheels and stabilizing the spacecraft, an image is taken in the \JE band for 112 s;
    \item the rotation of the filter wheel is repeated to obtain images in the \HE and\YE bands, for 112 s in each band.
\end{itemize}

This sequence is repeated four times to obtain dithered images and spectra dispersed with different orientations, with offset angles of $4\degree$. Taking into account also the dither slews and stabilization times, a ROS lasts 4214 s in total. 

Within each group of four dithers, the relative pointing offsets follow a fixed S-shaped layout on the sky (hereafter `S-sequence'), which consists of three relative offsets between the nominal pointing positions of consecutive dithers \citep{Markovic2017}. In focal-plane coordinates aligned with the detector axes, these angular offsets for DR1 can be expressed as
\begin{align}
(\Delta x,\,\Delta y) \in \left\{ (61'',\,111''),\,(0'',\,111''),\,(61'',\,111'') \right\} ,
\end{align}
where the first pair corresponds to the displacement between the first and second dither, and so on. The dominant shift is along the $y$ direction ($111''$), while the alternating $x$ component ($61''$) produces the characteristic S-shaped pattern across the full ROS block.
The orientation of the S-sequence is defined in the detector reference frame and is fixed for all ROS blocks in the EWS (Euclid Collaboration: Terreno et al, in prep.). 
In addition, the NISP grism dispersion axis rotates by $4^\circ$ between consecutive dithers.
The combination of the S-sequence offsets and the grism rotation generates a predictable angular redistribution of grism spectra on the focal plane. This geometry sets the characteristic angular scales at which persistence signals from bright sources are expected to recur across multiple exposures, and it forms the basis for the angular statistical diagnostics developed in this work.

\section{\label{sc:data}Catalogues}

\begin{figure}
    \centering
    \includegraphics[width=\linewidth]{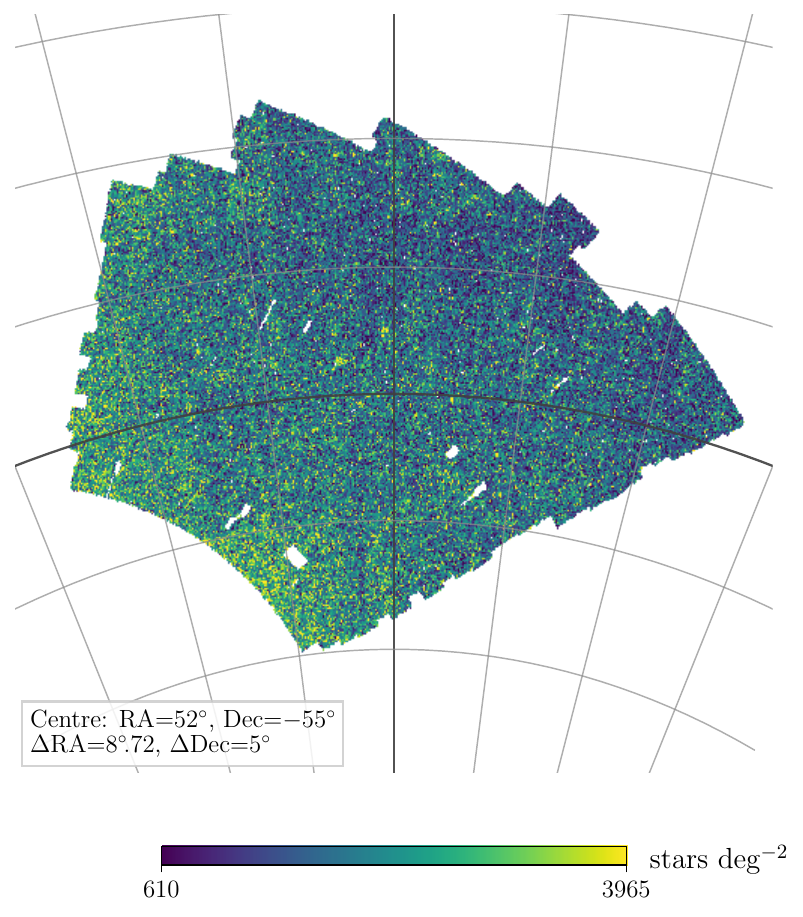}
    \caption{Location of the S1 field and \twomass\ star density inside its footprint.}
    \label{fig:S1_footprint_stardensity}
\end{figure}

In this work we made use of both real data and mock catalogues. Star samples were extracted from \Gaia\ \citep{Gaia_2016A&A595A1G, Gaia_2023} and \twomass\ \citep{2mass_2006AJ....131.1163S} catalogues, depending on the specific test performed. To simulate the spectroscopic galaxy sample, we employed the ELM \citep{EP-Monaco1}, which mimic the main observational properties of the \Euclid spectroscopic sample. In addition, we used the Flagship galaxy catalogue \citep{EuclidSkyFlagship} to perform preliminary tests aimed at modelling persistence effects. 
%(see Appendix~\ref{sc:FS2}).

To trade off between computational cost and statistical significance, we restricted our test to the S1 area in the southern Galactic hemisphere. Data in this region became available before the processing of the whole DR1 field and, for this reason, they have been extensively investigated during the preliminary galaxy clustering analysis in view of DR1. In Fig.~\ref{fig:S1_footprint_stardensity}, we show the location of the S1 field and the \twomass\ stellar density inside its footprint. It has an area of about $330 \sqdeg$.
%$329 \sqdeg$, and it is uniquely identified by the set of PatchId $[46, \,53, \,61]$ in the EWS. 

\subsection{Mock galaxy catalogues}

As starting point for both systematic effects analysed in this work, we used a subset of 100 mock catalogues extracted from the ELM suite. To match the characteristics expected for the \Euclid sample, \ha\ emitting galaxies were simulated by populating dark matter halos with galaxies using an halo occupation distribution (HOD) model calibrated on the Flagship mock galaxy catalogue. This has been constructed to reproduce the number density of \ha\ galaxies as given by model 3 of \cite{Pozzetti2016}. In this work, we used the `target' version of ELM, with galaxies selected above a fiducial \ha\ line flux threshold of $2\times10^{-16} \,\mathrm{erg}\,\mathrm{cm}^{-2}\,\mathrm{s}^{-1}$.
These catalogues contain \ha\ emitters above the target line flux threshold and are free from any additional selection effects; that is, they include neither redshift interlopers nor angular systematics arising from the dithering strategy or from the gaps between the 16 NISP detectors.
The original mock catalogues with a $30\degree$ radius light cone footprint have been centred on the S1 field and trimmed to match the 
% S1 footprintusing a healpix pixelization with \textcolor{magenta}{$NSIDE=8192(4096)$}\footnote{\url{https://euclid.roe.ac.uk/projects/gcswg/wiki/CreateSurveyFootprint}} of
area covered by the \Euclid pointings used to observe S1. Each catalogue is split into four redshift bins identified by their boundaries $\left\{0.9, 1.1, 1.3, 1.5, 1.8\right\}$. For each redshift bin, a single random catalogue is provided, containing 50 times more objects than the corresponding mock galaxy sample. The random objects are unclustered, with a uniform number density over the survey footprint and the same radial distribution as the simulated galaxies.

Without any loss of generality, we focused on the closest redshift bin of the \Euclid pre-launch baseline redshift intervals for the \Euclid galaxy clustering analysis, that is $z \in \left[0.9, 1.1\right]$. Table \ref{tab:mock-properties} reports the main properties of the ELM used in this work.

\begin{table}[h!]
\caption{Properties of the S1 target ELM used in this work.}
\label{tab:mock-properties}
\centering
\begin{tabular}{c c c}
\hline\hline
n° mocks & mean n° sources & redshift range \\
\hline
\noalign{\vskip 3pt}
100 & \num{237841} & $0.9 < z < 1.1$ \\
\noalign{\vskip 2pt}
\hline
\end{tabular}
\end{table}

When studying the effect of persistence on spectroscopic measurements, we made use also of the Flagship galaxy catalogue. Since its purpose is to provide a mock reference photometric sample, we considered only mock galaxies brighter than $\HE<24$.

\subsection{Star catalogues}

Depending on the type of analysis, we have considered both the \Gaia\ and the \twomass\ survey.

For the study on the star mask, we referred to the \Gaia\ catalogue, since it is the reference one used by MER. In particular, we used the third data release as in \citet{Q1-TP004}, downloading the corresponding catalogue from {\tt CosmoHub}.
Objects were selected by specifying the equatorial coordinates' limits of the S1 region and, to exclude extended sources, we applied the star–galaxy separation criterion used by the MER PF, requiring {\tt CLASSPROB\_DSC\_COMBMOD\_GALAXY < 0.1}. In our analysis, each \Gaia\ star is characterized by its {\tt source\_id}, its angular position in equatorial coordinates, and its {\tt phot\_rp\_mean\_mag} magnitude ($G_{\mathrm{RP}}$).

We considered \twomass\ data when studying persistence, since the magnitudes in the corresponding photometric bands are more representative of what we expect to measure with \Euclid near-infrared photometry. In particular, we made use of the Point Source Catalog (PSC), selected within the equatorial coordinates' limits of the S1 region. 
%\textcolor{magenta}{I didn't apply any further selection to exclude galaxies because there seems not to be a reliable criterium. Since the star--star correlation looks similar to Gaia, it may be that the same contamination is in both catalogues. Moreover, I found out that being far away from the galactic plane reduces the probability of misidentifying a galaxy for a star, which should be our case. See `Unresolved galaxies' section in \url{https://www.ipac.caltech.edu/2mass/releases/allsky/doc/sec2_2b.html?utm_source=chatgpt.com}}.
Table \ref{tab:stellar-properties} reports the properties of the star catalogues used in this work.

\begin{table}[h!]
\caption{Properties of the stellar catalogues used in this work, within the S1 footprint.}
\label{tab:stellar-properties}
\centering
\begin{tabular}{l c c}
\hline\hline
survey & n° sources & magnitude range \\
\hline
\noalign{\vskip 3pt}
\Gaia\ DR3   & \num{1470320} & $3.60 < G_{\mathrm{RP}} < 23.96$ \\
\noalign{\vskip 3pt}
\twomass\ PSC  & \num{673350}  & $1.63 < J_{\mathrm{2MASS,Vega}} < 18.98$ \\
           &         & $0.71 < H_{\mathrm{2MASS,Vega}} < 18.07$ \\
\noalign{\vskip 2pt}
\hline
\end{tabular}
\end{table}

\section{\label{sc:tools}Tools}

\subsection{\label{sc:2ptstat}Angular two-point statistics}

The main statistical tool used in this work is the angular two-point correlation estimator. We used it to search for and quantify possible stellar contamination in the \Euclid spectroscopic galaxy catalogue. In the absence of systematics, no intrinsic correlation is expected between foreground stars and spectroscopic galaxies. Therefore, any statistically significant star–galaxy cross-correlation signal can be interpreted as evidence of residual systematic effects

More specifically, our analysis relies on three angular correlation statistics: the star–galaxy correlation, used to identify potential stellar contamination; the galaxy–galaxy correlation function, used as reference when considering target galaxies; and the star–star correlation function, used to characterise the intrinsic angular correlation properties of the stellar sample and to assess the impact of possible contamination associated with extended stellar systems, such as star clusters.
To measure these statistics, we used the Landy--Szalay (LS) estimator \citep{LandySzalay1993} as implemented in the {\tt corrfunc} publicly available code \citep{corrfunc},
\begin{align}
    \textit{w}(\theta) = \dfrac{f_1 f_2 \, \mathrm{D}_1 \mathrm{D}_2 (\theta) - f_1 \, \mathrm{D}_1 \mathrm{R}_2(\theta) - f_2 \, \mathrm{D}_2 \mathrm{R}_1(\theta) + \mathrm{R}_1 \mathrm{R}_2(\theta)}{\mathrm{R}_1 \mathrm{R}_2(\theta)} \, .
\end{align}
Here, the indices 1 and 2 identify the type of sample, indicating whether the objects are stars or galaxies. The quantity $f_{i }=N_{R_{i}}/N_{D_{i}}$ represents the ratio between the number of sources of type \textit{i} in the random catalogue and the number of data sources of type \textit{i}, where \textit{i} can be either 1 or 2.

We expect angular systematics due to both star mask and persistence to leave a signature detectable at small scales in the star--galaxy cross-correlation. Any mismatch between data and random in the stellar mask will be visible at the masking scales, which correspond to at most $40 \arcsec$ in \Euclid (as detailed in Sect.~\ref{sec:mermasking}). Similarly, we expect to see persistence effects at the dithering scale. 
Therefore, we estimated the angular correlation functions over a sufficiently large range of scales, using a binning fine enough to capture possible signatures of star-related effects.
We computed angular separations in 30 logarithmically spaced bins spanning $1\arcsec$ (set by the smallest masking scale of the faintest stars) to $1\degree$, well beyond the angular imprint of the dithering pattern.
For persistence, which affects also angular scales comparable to those of the ROS, we extended the analysis to separations up to $15\degree$ in linearly spaced bins of width $\ang{0.1}$. This allowed us to capture potential periodic replicas of the persistence pattern and, at the same time, to probe scales well beyond the BAO scale.

To compute the uncertainty on the angular two-point correlation functions, we must distinguish between the different types of statistics. Since we have 100 realizations for the galaxies in the mock catalogues, we computed the uncertainty on the star--galaxy and galaxy--galaxy correlations as the scatter among the realizations. For the stars' auto-correlation, we accounted for the Poisson error associated with the star pair counts.

\subsection{2PCF}

In order to estimate the 2PCF, we made use of the methodology and software developed within the \Euclid Science Ground Segment \citep{EP-delaTorre}. The latter utilizes the minimum-variance LS estimator and enables the use of the random split method \citep{ElinaKeihanen2019} to speed up the computation. We evaluated the monopole and the quadrupole moments of the anisotropic 2PCF  using \num{40} equally spaced bins in separation $r \in [0,200]\, \hMpc$ ($\Delta r = 5\, \hMpc$) and \num{100} bins in $\mu$ within $\mu \in [-1,1]$, where $\mu$ is the cosine of the angle between the galaxy-pair separation vector and the line of sight. To compute the comoving separations of galaxy pairs from their redshifts, we adopted the same cosmological model used to generate the ELM catalogues. 
%\sout{Without any loss of generality, we focused on closest redshift bin of the \Euclid pre-launch baseline redshift intervals for the \Euclid galaxy clustering analysis, that is $z \in \left[0.9, 1.1\right]$.}

\subsection{Angular metric}

Deviations of the angular star--galaxy cross-correlation from zero indicate the presence of some systematic effects, related to an imperfect treatment of the star components, affecting the galaxy catalogue. The magnitude of this effect is expected to depend on the intrinsic angular correlation of the galaxies. To quantify the systematic effect and compare cross-correlations measured in different configurations, we therefore introduce the normalised star--galaxy cross-correlation
\begin{align}
    \mathrm{SG} (\theta) = \left\langle \dfrac{w_{\mathrm{sg}}(\theta)}{|w_{\mathrm{gg}}(\theta)|}\right\rangle \, ,
\end{align}
where the average runs over all mock galaxy realizations.
To quantify the statistical significance of deviations of the star--galaxy signal from zero, we use the metric
\begin{align}
\label{eq:angular-metric}
    \mathrm{metric} (\theta) = \dfrac{\mathrm{SG} (\theta)}{\sigma_{\mathrm{SG}}} \, ,
\end{align}
where $\sigma_{\mathrm{SG}}$ is the the scatter associated the normalised star--galaxy cross-correlation measured over the set of mock catalogues.

\subsection{\label{sc:gelsa}{\tt GELSA} simulator}

To simulate the persistence effect, it is necessary to compute the position of sources on the NISP focal plane in photometric and spectroscopic exposures. We used the {\tt GELSA} library for this purpose (Euclid Collaboration: Granett et al, in prep.).
Developed as part of the ELSA\footnote{\url{https://elsa-euclid.github.io/\#}} project (Euclid Legacy Science Advanced analysis tools), {\tt GELSA} reads the astrometric and dispersion solutions from the NIR and SIR pipelines and computes coordinate transformations \citep{Q1-TP003, Q1-TP006}. As described in Sect.~\ref{sc:simulating-persistence}, we used the code to compute the detector pixel coordinates of stars in order to locate the persistence images. Then, given a spectroscopic exposure, we transformed from pixel coordinates to equatorial coordinates to assign the interlopers' position in the sky.

% (Gabarra et al., in prep).

% \code{gelsa} version: `main' branch, right after 42ea7d1e3785addf7a653b286dab9fccd310cdc4 commit SHA. 

% Calibration files and configuration files from \code{gelsa-spectra/config}, last commit: c163356e9ebecbfa548c24d97ee2d04317fa3e08 commit SHA.

%%%%%%%%%%%%%%%%%%%%%%%%%%%%%%%%%%%%%%%%%%%%%%%%%%%%%%%%%%%%

\section{\label{sc:persistence}Modelling and measuring photometric persistence in spectroscopic images}

%\subsection{Context and simplified simulation procedure}

Photo-electrons collected in the NISP detector pixels are read out and, in principle, reset before the subsequent exposure in the observing sequence. In practice, this reset is not perfect. Due to charge traps in the semiconductor material, a fraction of the photo-electrons remains temporarily stored in metastable states and is released only at later times. This delayed release produces a spurious signal that decays on timescales significantly longer than the interval between consecutive exposures \citep{persistence-theory}.
%In this work, we assessed the impact of this effect on galaxy-clustering measurements by adopting a simplified model of persistence from photometric to spectroscopic exposures (hereafter `photo-to-spectro' persistence).
Imperfect removal of this residual signal can mimic emission features in spectroscopic frames, ultimately leading to the inclusion of spurious detections in the galaxy catalogue, commonly referred to as interlopers.

To illustrate the mechanism, consider the image of a bright star recorded by the NISP detector during a photometric exposure. If the accumulated charge is not completely removed during the reset phase, a residual imprint of the star may persist in the same detector pixels and contaminate the subsequent spectroscopic exposure. Although the amplitude of this residual signal decreases with time, its integrated contribution over the full spectroscopic exposure can still exceed the noise level.
If the residual signal spatially overlaps with the dispersed spectrum of a real galaxy, it may mimic an emission feature in the spectrogram. This can lead to an incorrect redshift assignment. In this scenario, the affected source is a genuine galaxy whose redshift has been catastrophically misestimated.

%In DR1, persistence will be mitigated by the NIR PF \citep[Near InfraRed imaging data,][]{Q1-TP003}, which masks pixels predicted to be significantly affected \citep{DR1-TP038}. Here, we investigate hypothetical residual persistence that could remain after this mitigation, for example because of simplifying assumptions in the adopted temporal or geometrical masking model or because low-level persistence is not explicitly masked. Our goal is not to estimate the residual contamination expected in DR1, but to characterise the statistical signature that such residuals would produce if present.Here, we focus on the residual persistence signal that may remain because of incomplete or imperfect masking.

The persistence of bright stars in spectroscopic exposures is expected to leave a characteristic signature in the angular clustering signal. Since the overlap between persistence residuals and galaxy spectra does not occur at random locations on the focal plane, but is instead tied to the detector positions where the stellar signal was previously recorded, the resulting contamination should exhibit a distinctive and predictable angular pattern.
The characteristic angular scales of this imprint are set by the combined effects of the dithering strategy, the rotation of the grism dispersion direction between consecutive exposures, and the intrinsic angular clustering of stars on the sky. Together, these elements determine both the scale dependence and the morphology of the induced spurious correlation, providing a direct diagnostic to identify and quantify its presence in the data.

In practice, several additional factors contribute to shaping the angular pattern induced by persistence. 
%First, the decay properties of the persistence signal is pixel-dependent and varies across the focal plane. 
%Moreover, the decay timescale depends on the residual flux at the end of the photometric exposure, which is directly related to the stellar magnitude. 
%As a consequence, brighter stars remain imprinted on the focal plane for a larger number of subsequent exposures than fainter ones.
First, the amplitude of the persistence signal and the coefficients describing its power-law decay are both pixel-dependent (they vary across the focal plane) and flux-dependent (they vary with the star magnitude). These coefficients have been measured over a broad range of signal levels in ground-based tests \citep{EU-Kubik} and, more recently, for a limited number of in-flight measurements \citep{EU-Kubik, Kubik:2024spie}. 
%In this work, we adopted a single set of coefficients corresponding to a reference initial flux level. 
%\textcolor{magenta}{the persistence model is still under refinement, and a single set of coefficients corresponding to a single initial flux level was provided and adopted for this work.}
Furthermore, the decay relation derived for non-saturated sources is not expected to hold exactly for saturated objects such as the brightest stars.
%In this work, we do not attempt to model all these effects in detail. A rigorous treatment of persistence would require a precise pixel-level characterization of the effect, which is not yet fully available, together with dedicated and computationally intensive end-to-end simulations. This is beyond the scope of the present analysis. Instead, we provide a first-order assessment of the impact of persistence on clustering measurements, adopting a simplified framework designed to capture the dominant effects while bypassing the full complexity outlined above. Our goal is to establish whether star-related persistence can affect clustering analyses and, if so, to identify the relevant angular scales and estimate the order of magnitude of the induced contamination.
A rigorous treatment of persistence would require a detailed pixel-level characterization of the detector response, including its dependence on illumination history, spatial variations across the focal plane, and non-linear decay behaviour, as discussed in Euclid Collaboration: Kubik et al. (in prep.).
%\citet{DR1-TP038}. 
Such an approach would also require dedicated end-to-end simulations and is beyond the scope of this work. Instead, in this work we provide a first-order, phenomenological assessment of the impact of persistence on clustering measurements, designed to capture the dominant effects while neglecting detector-level complexities.

\subsection{Persistence model}
\label{sc:persistence-model}

The persistence model adopted for the NISP detectors is described in \cite{EU-Kubik}. In that work, the temporal decay of the persistence signal is shown to follow a power-law behaviour, with parameters that depend on the initial flux level of the source.
In this analysis, we are interested in the persistence signal integrated over the duration of the subsequent spectroscopic exposure. This integrated charge must be compared to the typical spectroscopic background level, which is $\sigma_{\mathrm{bkg}} \simeq 30 \, e^-$ for a typical spectroscopic exposure \citep{Q1-TP006}.
%\sqrt{1000} --> 30

To compute the persistence charge $P$ accumulated during a spectroscopic exposure, the \Euclid NIR team adopts the expression\footnote{This equation corresponds to the integral of equation 2 in 
%\citet{DR1-TP038}
Euclid Collaboration: Kubik et al. (in prep.).}
\begin{align}
\label{eq:pers-current}
P = I_0 \frac{\rm OFFSET}{\rm SLOPE} \left(t_2^{\rm SLOPE} - t_1^{\rm SLOPE}\right) \, ,
\end{align}
where $I_0$ denotes the number of electrons measured at the end of the photometric exposure preceding the spectroscopic frame under consideration, and $t_1$ and $t_2$ correspond to the start and end times of the spectroscopic exposure, measured with respect to the midpoint of the photometric exposure that generated the persistence signal. 
The parameters $\rm OFFSET$ and $\rm SLOPE$ depend on the detector and pixel, as well as on the incident flux \citep{Kubik:2024spie}. 
%In this work, we adopted a single set of coefficients corresponding to a reference initial flux level. 
As a first approximation, we use a model calibrated for a reference $I_0=30 \, \mathrm{k}e^-$ and use the median values of the $\mathrm{OFFSET}=0.0064$ and $\mathrm{SLOPE}=-0.0734$, computed after masking bad pixels, for all pixels. 

% OFFSET and SLOPE matrices are specified for each pixel of the 16 NISP detectors here: \footnote{\code{EUC\_NIR\_Persistence\_PowerLaw\_Model\_001\_241213.fits}}. \textcolor{gray}{Why do we have only these coefficients in the DPS and not the equivalent for a list of incident fluxes like in \citealt{Kubik:2024spie}?}

% Bad pixels \footnote{\code{EUC\_NIR\_C-BADPIXEL\_MASK\_20250408T170633.544398Z.fits}}. 

The detailed procedure used to convert stellar magnitudes into detector-level electron counts and to estimate the corresponding values of $I_0$ is described in Appendix \ref{app:from-mag-to-ne} and \ref{app:I0modelling}, respectively. We adopted a simplified treatment of the stellar signal, approximating the persistence contribution using the flux in the central pixel of the star image and neglecting its full two-dimensional extent. This approximation is sufficient for our purposes, as we are primarily interested in capturing the order of magnitude and spatial distribution of the persistence-induced contamination.

\subsubsection{Persistence duration and effective number of affected exposures}

\begin{figure}
    \centering
    \includegraphics[width=\linewidth]{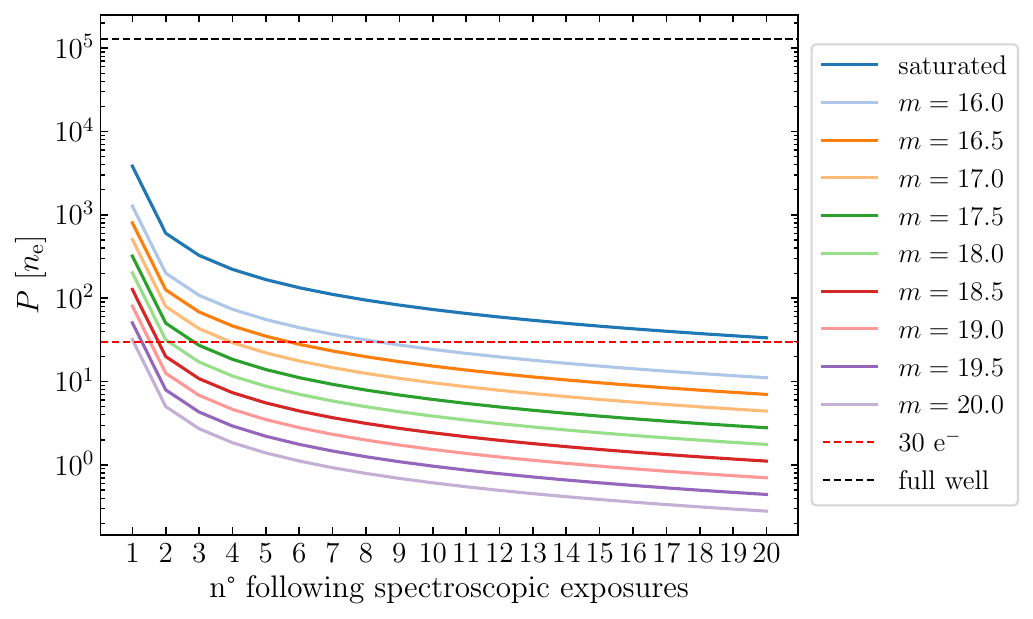}
    \caption{Persistence charge in the spectroscopic exposures following the initial photometric \YE exposure. Different colours represent different star magnitudes. The dashed black line indicates the full well of the NISP pixels (Appendix~\ref{app:I0modelling}), whereas the red dashed line indicates the typical background level of a \Euclid spectroscopic exposure.} 
    \label{fig:ne-vs-following-nexp}
\end{figure}

%In principle, the persistence model should be applied to every star in the catalogue, accounting for the fact that, depending on its magnitude, the residual signal may propagate into a different number of subsequent spectroscopic exposures. Implementing this approach for a realistic sequence of ordered \Euclid pointings would be computationally expensive.
%However, since no intrinsic spatial correlation is expected between stars and target galaxies, we adopted a simplified strategy. 
%Rather than tracking the exact number of affected spectroscopic exposures for each individual star, we introduced an effective number of spectroscopic exposures, $\neff$, assumed to be common to all stars irrespective of their magnitude.
The persistence model should be applied to every star in the catalogue, taking into account that the number of affected exposures depends on its brightness. However, modelling this dependence for the complex sequence of \Euclid exposures would be computationally expensive. Therefore, and given that no angular correlation is expected between galaxies and stars, 
we neglected this dependence and we adopted a simplified strategy in which we considered a single effective number of exposures $\neff$ affected by persistence, irrespective of the star brightness.
In the following, we describe the procedure used to estimate this effective number of spectroscopic exposures.

Based on the \Euclid \YE magnitudes of the \twomass\ sample (computed following Appendix \ref{app:from-mag-to-ne}), which span the range $2.5 \leq m \leq 20.3$, we considered a magnitude interval $2 < m < 20.5$ divided into bins of width $\Delta m = 0.5$. For the central magnitude of each bin $m_i$, we computed the number of subsequent spectroscopic exposures $N_{\rm exp}(m_i)$ for which the persistence signal is expected to remain above a background noise level  $\sigma_{\mathrm{bkg}} = 30 \, \mathrm{e}^-$.
To retrieve $N_{\rm exp}(m_i)$, we needed to compute the persistence signal using Eq.~\eqref{eq:pers-current}, and therefore we needed to specify the time elapsed between the photometric exposure and the following spectroscopic exposures, as well as the duration of both photometric and spectroscopic exposures. We adopted the parameters (based on Fig.~\ref{fig:surveysequence})
\begin{align}
t^{\mathrm{photo}} &= 112 \,\mathrm{s} \, ,\\
t^{\mathrm{spec}} &= 574 \,\mathrm{s} \, ,\\
t_0 &= 10 \,\mathrm{s} \, ,\\
\Delta t ^{\mathrm{spec}} &= 1066 \,\mathrm{s} \, .
\end{align}
Here $t^{\mathrm{photo}}$ and $t^{\mathrm{spec}}$ denote the exposures time of the photometric and spectroscopic observations, respectively. The parameter $t_0$ represents the time interval between the end of the photometric exposure and the beginning of the subsequent spectroscopic one. The quantity $\Delta t^{\mathrm{spec}}$ corresponds to the time separation between consecutive spectroscopic exposures, including both the acquisition time of the photometric and spectroscopic frames (about $900 \, \mathrm{s}$) and the average slew time between dithers ($70 \, \mathrm{s}$). 
Referring to Eq.~\eqref{eq:pers-current}, the value of $I_0$ is computed as described in Appendix \ref{app:I0modelling}. For the $n^{\rm th}$ subsequent spectroscopic exposure, the integration limits of the persistence signal are therefore
\begin{align}
t_1 &= \dfrac{t^{\mathrm{photo}}}{2} + t_0 + n\,\Delta t ^{\mathrm{spec}} \, ,\\
t_2 &= \dfrac{t^{\mathrm{photo}}}{2} + t_0 + t^{\mathrm{spec}} + n\,\Delta t ^{\mathrm{spec}} \, .
\end{align}
The resulting persistence charge for the first 20 subsequent spectroscopic exposures is shown in Fig.~\ref{fig:ne-vs-following-nexp} for different stellar magnitudes (and therefore for different initial values of $I_0$ in Eq.~\ref{eq:pers-current}).
The number of exposures $N_{\rm exp} (m_{\rm i})$ with a significant persistence signal is found in correspondence of the crossing of each curve with the background noise level (horizontal red dashed line). Notably, even relatively faint stars with magnitudes $m \in [19,20]$ are expected to produce a detectable persistence signal in the immediately following spectroscopic exposure.

Then, we divided the \twomass\ stars within the S1 footprint into the same magnitude bins used to compute $N_{\rm exp}(m_i)$ and counted the number of stars $N_{\rm stars}(m_i)$ in each bin. 
Finally, we estimated the effective number of persistence-affected exposures as a weighed average
\begin{align}
   \neff = \dfrac{\sum_{\rm i} N_{\rm exp} (m_{\rm i}) N_{\mathrm{stars}}(m_{\rm i})}{\sum_{\rm i} N_{\mathrm{stars}}(m_{\rm i})} \, ,
\end{align}
where each value $N_{\rm exp}(m_i)$ is weighted by the number of stars in the corresponding magnitude bin.
In this way, the most abundant stars contribute more strongly to the estimate of $\neff$ than the relatively rare very bright stars, which would otherwise lead to a disproportionately large number of affected exposures. The resulting value is $\neff = 8$, meaning that, on average, the persistence signal from a photometric exposure acquired in a given dither remains significant for the following eight spectroscopic exposures.
%\textcolor{magenta}{(with $3 \times fw$ it's 8, with 2 it was 7. However, the feature at small scales doesn't change position neither with much more significant changes}.
This value should be regarded as conservative, since it overestimates the persistence effect.  In fact, some S1 pointings are separated by time intervals larger than $\Delta t^{\mathrm{spec}} = 1066 \,\mathrm{s}$ due to larger slews between different quadruplets of dithers. In such cases, the longer time separation would allow the persistence signal to decay further, leading to a slightly smaller number of spectroscopic exposures affected by persistence in those regions of the sky.

\subsection{Modelling persistence contamination in mock catalogues}
\label{sc:simulating-persistence}

Although we can predict the number of persistence images from bright stars, it is not possible to directly estimate the level of contamination in the selected redshift catalogue. This is because the contamination depends on the effectiveness of masking in the spectroscopic extraction pipeline, and the selection criteria used to build the sample.

We can build an intuition on the severity of the contamination by considering the number of persistence images that fall on an extracted spectrum and potentially lead to a wrong redshift determination.
We used the \Euclid Flagship simulation to investigate this point, since it contains the full $\HE < 24$ flux-limited photometric sample.
As shown in Appendix~\ref{sc:FS2}, 66\% of persistence images fall on an extracted spectrum. Of these contaminated spectra, 97\% are part of the photometric sample, and not target ELGs. It is thus very likely that a persistent signal will overlap an extracted spectrum of the photometric galaxy sample and be interpreted as an emission line, leading to a spurious ELG detection.
We further assume that persistence signals overlap exclusively with sources that are not part of the target ELG sample, such that new spurious sources with a wrong redshift should be added to the ELG mock catalogue. 

We developed the following procedure to add interlopers due to persistence in the mock catalogue. 
We simulated all persistence images generated by the observation of stars in the S1 region along the sequence of \Euclid pointings in that field. For each pointing, we tracked the position on the focal plane of persistence signals produced by stars observed in the previous $\neff = 8$ pointings.  
For each pointing, we converted the focal plane coordinates of the persistence images into sky coordinates corresponding to the position of a hypothetical photometric galaxy. To do so, we assumed that each persistence signal was misidentified as an \ha\ emission line at a random wavelength within the NISP detection range. Under this assumption, each persistence image was associated with the sky position that a galaxy would have if such a line were observed at that focal-plane location in that specific pointing.
Because the same star can generate persistence signals in several subsequent exposures, a single star can promote several photometric galaxies to interlopers that contaminate the spectroscopic catalogue.
The corresponding sky positions differ from exposure to exposure, since the mapping from focal plane coordinates to angular coordinates depends on the centre of the pointing, as well as on the grism dispersion direction and the wavelength assigned to the persistence signal in that specific dither.

Finally, among all possible overlaps generated in this way, we randomly selected a subset of sources such that they represent a fraction $f$ of the total observed sample (galaxies plus persistence contaminants). These sources were then added to each mock catalogue of spectroscopic galaxies, thus producing a set of 100 contaminated mock catalogues with a $f$ contamination level. 

The effect of persistence contamination is similar but not entirely analogous to the one of `noise interlopers' described by \cite{EP-Risso}, because in the first case the persistence contamination imprints an angular clustering signal that is significantly stronger than the intrinsic clustering of the photometric galaxies. The latter is diluted by projection effects within the observed redshift bin. This justifies our choice of randomly subsampling the modelled persistence signal to the desired contamination level, thus neglecting the intrinsic clustering of the photometric galaxies which become interlopers due to the persistence contamination. 

\subsubsection{Setting up the angular correlation analysis for persistence}
\label{sc:setup-angcorr-persistence}

We tested different fractions of contamination due to persistence, ranging from 10\% to 50\%, with $f=50\%$ representing an extreme scenario. We expect the contamination fraction to vary with Galactic latitude, as it correlates with the stellar density, and with the density of spectra extracted through the spectroscopic pipeline. However, the contamination fractions presented here should be regarded as controlled phenomenological scenarios designed to characterise the statistical signatures of persistence, rather than as predictions for the level of residual contamination expected in DR1.

To identify and characterize persistence contamination, we computed the angular star--galaxy, galaxy--galaxy, and star--star two-point correlations. Since persistence is expected to strongly affect the angular distribution of the sample rather than its radial one (we expect a random redshift distribution for persistence contaminants, like noise interlopers in \citealt{EP-Risso}), we restricted the analysis to the reference redshift bin $z \in [0.9,1.1]$, where the galaxy number density is highest.

For each contamination fraction, we computed the angular auto- and cross-correlation functions for every mock catalogue and then evaluated their averages over the full ensemble of mocks described in Sect.~\ref{sc:simulating-persistence}. The stellar catalogue was constructed from the original rectangular \twomass\ catalogue covering the S1 region by restricting it to the S1 footprint, so as to match the angular selection of the mock catalogues. This catalogue was then used for all star--galaxy and star--star measurements presented in this section.

We used {\tt corrfunc} to compute all angular correlation functions. First of all, these were evaluated in 30 logarithmically spaced bins spanning the range $\theta \in [1\arcsec, 1\degree]$. Then, we extended the analysis to separations up to $15\degree$ in linearly spaced bins of width $\ang{0.1}$. 
We used the same random catalogue for the mock galaxy catalogue and star catalogue. When extending the analysis to large angular separations, we adopted a downsampled version of the random catalogue and implemented the random split method \citep{ElinaKeihanen2019} for both the galaxy--galaxy and star--galaxy correlations.

% **COMMENTED BY BEN** 
%The same random catalogue homogeneously distributed across the S1 area, namely the original random catalogue of the uncontaminated mocks, was used for both galaxies and stars. This choice is motivated by the absence of angular selection effects in both the parent (i.e. uncontaminated) and contaminated mock galaxy catalogues, and by the small variation in stellar surface density across the relatively compact S1 footprint. The only consequence of using a common random catalogue is a suboptimal shot-noise correction, since the ratio of data to random objects differs between the star and galaxy catalogues and between galaxy catalogues with different contamination fractions; however, this effect is expected to be minor. When extending the analysis to large angular separations, we adopted a downsampled version of the random catalogue and implemented the random split method \citep{ElinaKeihanen2019} for both the galaxy--galaxy and star--galaxy correlations.
%in order to keep the computational cost manageable.

All analyses presented in this section refer to quantities that are directly measurable in the EWS. In Appendix~\ref{app:nonmeascorrelations-persistence}, we also discuss angular correlation statistics involving persistence-induced interlopers, such as the persistence--persistence and star--persistence correlations. Although these quantities are not directly measurable in the EWS and could only be isolated in the Euclid Deep Fields, they are useful to interpret the behaviour of the observable statistics discussed in the main text.

\subsection{Results on persistence: small scales}
\label{results-persistence-ss}

\begin{figure*}
    \centering
    \includegraphics[width=0.33\linewidth]{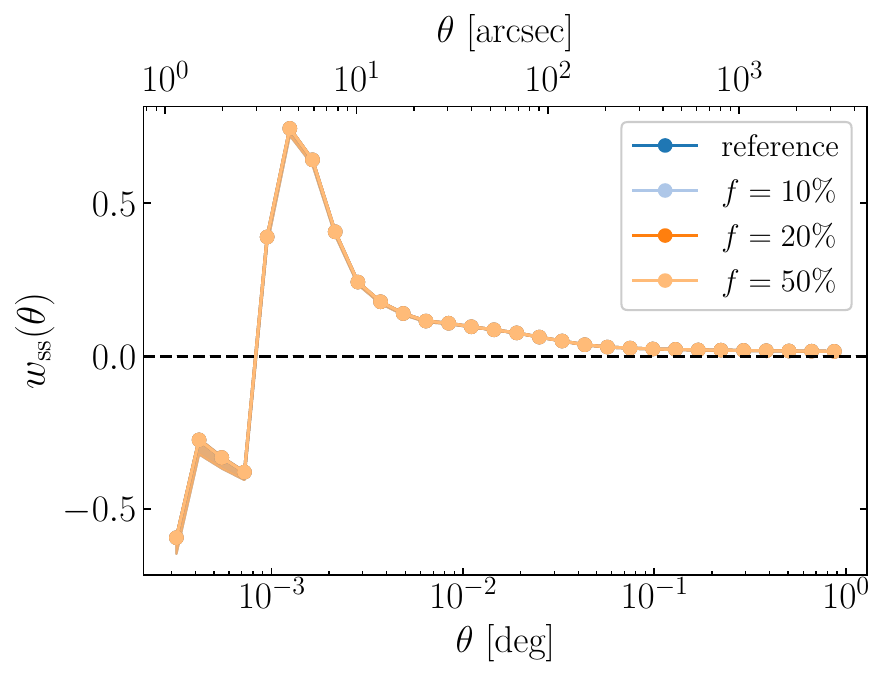} %\hfill
    \includegraphics[width=0.33\linewidth]{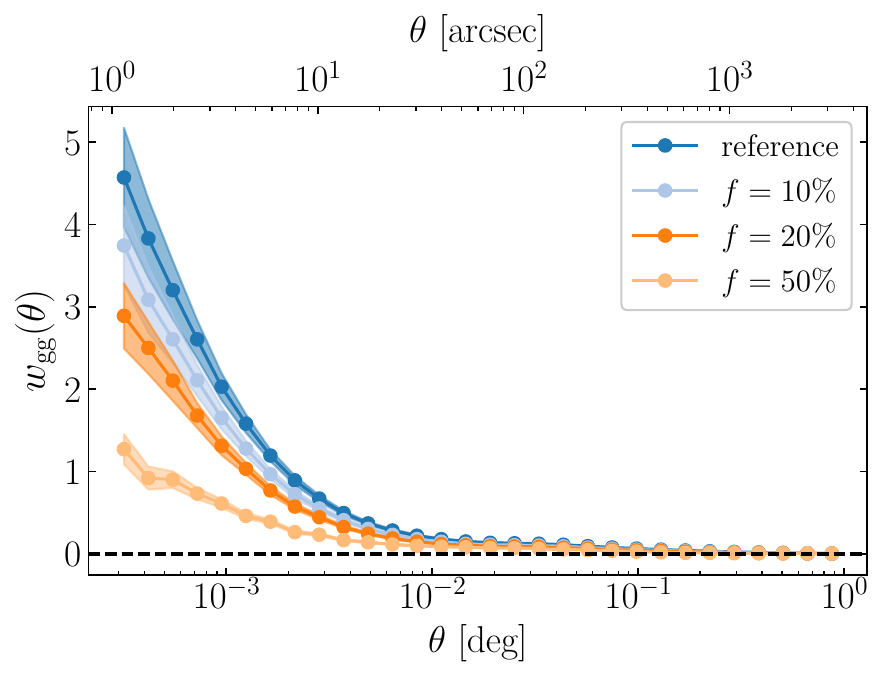} %\hfill
    \includegraphics[width=0.33\linewidth]{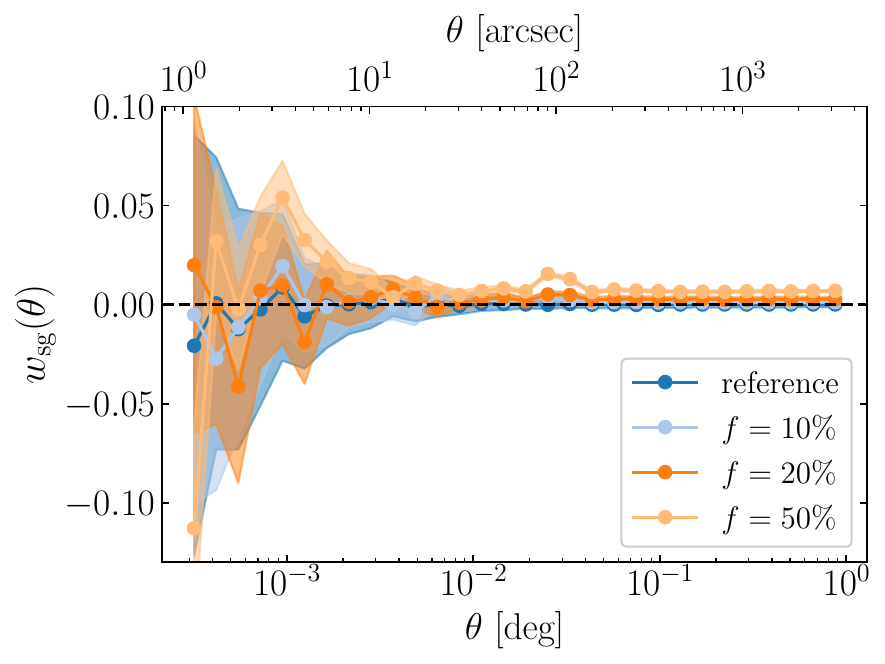}
    \caption{Angular statistics averaged over the 100 mocks for the reference uncontaminated case (blue line) and for the persistence-contaminated scenarios. \textit{Left}: star--star angular auto-correlation (which is independent on the contamination of the galaxy sample). \textit{Centre}: galaxy--galaxy angular auto-correlation. \textit{Right}: star--galaxy angular cross-correlation. The shaded areas represent the scatter over the mock measurements for $w_{\mathrm{gg}}$ and $w_{\mathrm{sg}}$, and Poisson errors for $w_{\mathrm{ss}}$.}
    \label{fig:ideal-correlations}
\end{figure*}

Figure \ref{fig:ideal-correlations} shows the star–star, $w_{\mathrm{ss}}$, galaxy–galaxy, $w_{\mathrm{gg}}$, and star–galaxy, $w_{\mathrm{sg}}$, angular correlation functions in the selected redshift bin. We display both the reference case (blue lines), corresponding to the case of no persistence contamination, and the contaminated cases with different contamination fractions. The error bars shown in the figure represent the scatter over the full set of 100 mock realisations. The only exception is the star–star correlation, for which the error bars correspond to Poisson uncertainties.
%, since only one stellar catalogue was used.

Let us first consider the star--star correlation. 
We find results qualitatively similar to those of \cite{starsSDSS}. There is a positive correlation at small angular separations, corresponding to the typical size of bound stellar systems such as binaries, followed by a monotonic decrease toward zero on larger scales. Here, the decrease is more gradual and shows a more complex structure due to the presence of the prominent globular cluster NGC~1261 \citep{NGC1261, Wenger2000}, whose angular size of about $7\arcmin$ is comparable to the scale of the non-zero feature visible in the star--star correlation. 
In Appendix~\ref{app:globclusS1} we show that the shape of the star--star correlation recovers the expected behaviour when the region containing the globular cluster is excluded. 
The turnover around $3\arcsec$ reflects the angular resolution of the star catalogue. 
%and is observed in both the \twomass\ and \Gaia\ samples (see Appendix \ref{app:gaiavs2mass}).

The galaxy–galaxy auto-correlations exhibit the typical featureless power law, with an amplitude that anti-correlates with the contamination fraction.
This is expected, as \cite{EP-Risso} have shown that the shape and amplitude of the galaxy--galaxy correlation results from the weighted sum of the contribution of correct galaxies, weighted by the the square of the purity of the sample, and of the auto-correlation of the fake sources due to persistence weighted by the contamination fraction (shown in Appendix \ref{app:nonmeascorrelations-persistence}).

Let us now focus on the star--galaxy angular cross-correlation. In the reference case, when no persistence sources are added to the mock galaxies, we expect this correlation to be consistent with zero at all angular scales. 
%\sout{In our measurements, however, we observe a small deviation from zero at very small separations (below $4\arcsec$), which follows the same trend as the anti-correlation observed in the star--star correlation.  This is due to the fact that, given that anti-correlation in the star catalogue, it is less likely to find star--galaxy pairs at those scales as well. }
Deviations from zero are
%\sout{instead} 
observed in the contaminated catalogues, the most significant of which is a correlation peak around $100\arcsec$, particularly evident in the case of extreme $f = 50\%$ contamination. This peak is driven by an excess of star–interloper pairs at separations ranging from about 
$70\arcsec$ to $260\arcsec$. The exact scale of the excess depends on the wavelength assigned to the persistence signal and on the tilt of the grism used in the spectroscopic exposure in which that signal was interpreted as an \ha\ line. When the wavelength assignment is randomized for each persistence image, these excess pairs contribute to a broad feature around 
$100\arcsec$. The presence of this feature should not be underestimated, as it provides a practical tool to detect persistence-induced contamination and to assess its magnitude.

\begin{figure}
    \centering
    \includegraphics[width=\linewidth]{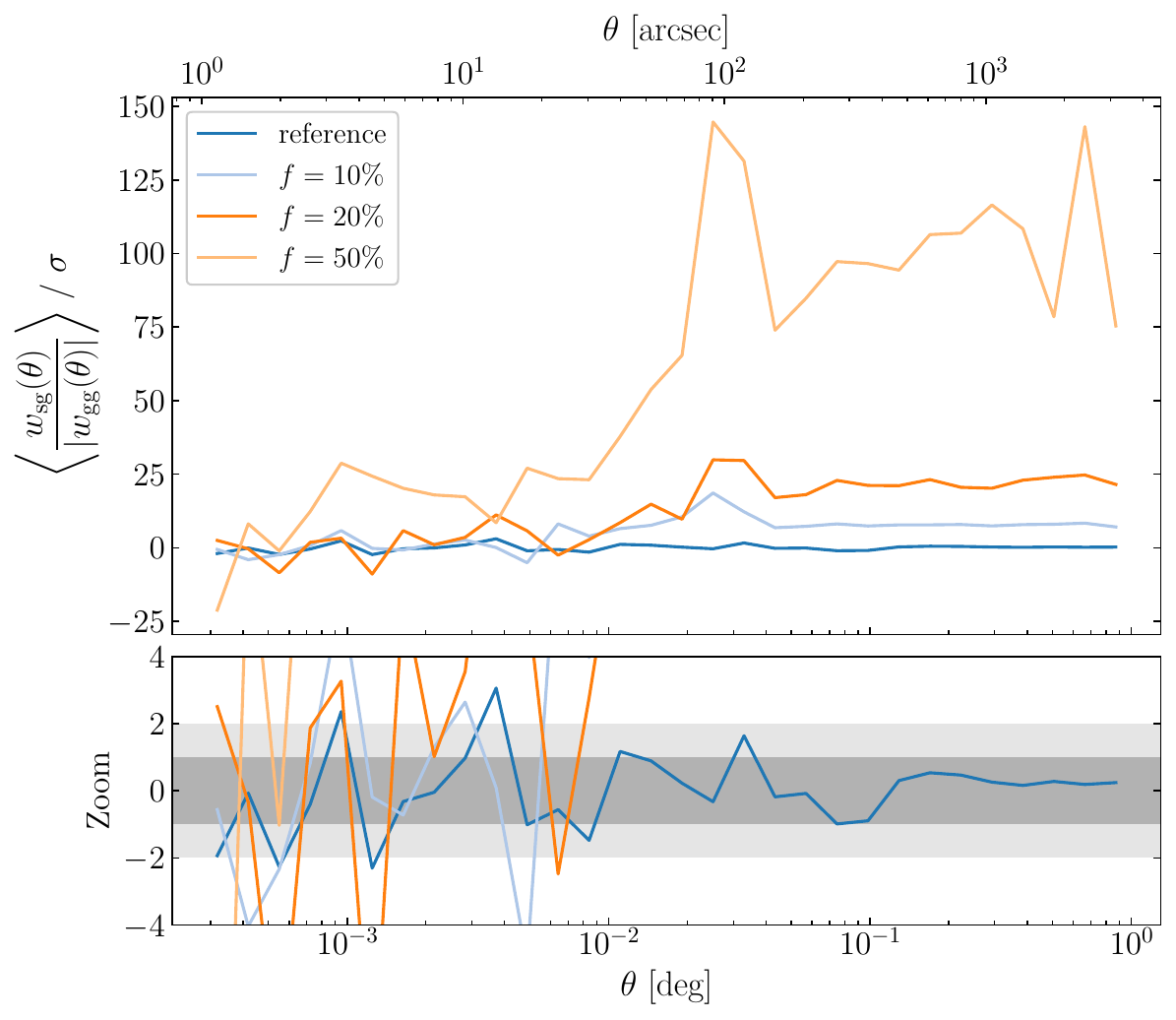} %\hfill
    \caption{Angular metric for the uncontaminated catalogues and the persistence-contaminated ones. Shaded grey areas refer to $1\sigma$ and $2\sigma$, where $\sigma$ is the error on the mean of the metric over all mocks. 
    %We show three cases corresponding to three different uncertainties: the error on the mean of the metric (\textit{top}), the error on the single measurement on S1 (\textit{bottom}, i.e the scatter on all measurements), and the error rescaled to the DR1 area (\textit{centre}).
    }
    \label{fig:persistence-metric-significance}
\end{figure}

The metric defined in Eq.~\eqref{eq:angular-metric} and shown in Fig.~\ref{fig:persistence-metric-significance} for all the cases explored allows us to assess the significance of deviations from the expected null cross-correlation. The excess peak clearly represents the most significant deviation. In the figure, the significance is computed with respect to the average over 100 realizations. For a single realization, corresponding to the analysis of the \Euclid S1 field, the significance would be approximately an order of magnitude smaller than that shown, yet still large enough to allow detection of the peak even in the case of 10\% contamination.

To estimate the detectability of such deviations in DR1, we rescaled the uncertainty measured for a single realization in S1 (which is the scatter among the different realizations of the metric) to the DR1 survey area, which is approximately $7.6$ times larger than the S1 footprint. Under DR1-like statistical uncertainties, the peak around $100\arcsec$ would be detectable with a significance greater than $4\sigma$ even for a contamination fraction of $f = 10\%$. This suggests that residual, unmasked persistence signals at this level or higher should be detectable in the Euclid DR1 data. 

For completeness, we report in Appendix \ref{app:nonmeascorrelations-persistence} the shape of the persistence--persistence signal computed at the same angular scales of the directly measurable correlations just showed.

\subsection{Results on persistence: large scales}
\label{results-persistence-ls}

\begin{figure}
    \centering
    \includegraphics[width=\linewidth]{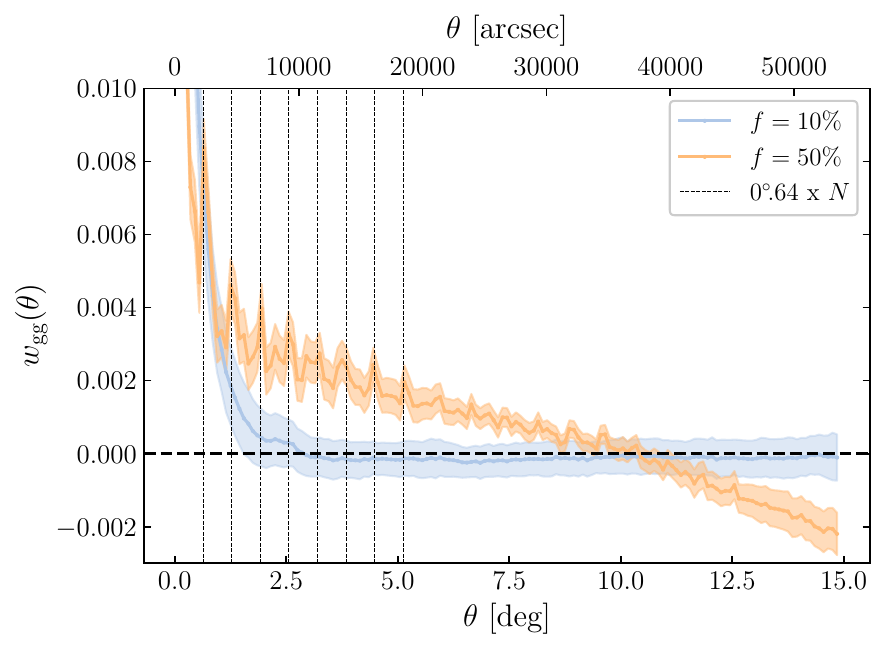} \hfill
    \includegraphics[width=\linewidth]{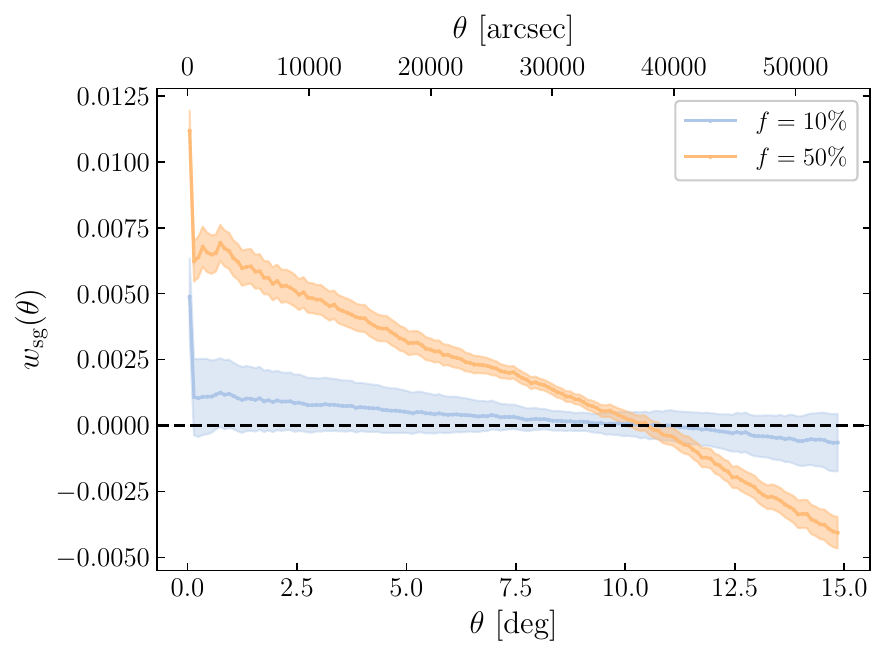}
    \caption{Galaxy--galaxy auto-correlation (\textit{top}) and star--galaxy cross-correlation (\textit{bottom})  at large scales for $f=10\%$ and $f=50\%$. The shaded areas represent the scatter over the mock measurements. Dashed vertical lines represent the first $N$ periodicities of the angular scale corresponding to the field of view size.}
    \label{fig:ls-angstat-f10f50}
\end{figure}

By construction, persistence signals are replicated following the sequence of \Euclid\ pointings and therefore reflect the step-and-stare observing strategy described in Sect.~\ref{sc:observation-strategy}, in which neighbouring regions of the sky are observed through four slightly dithered pointings. As a consequence, persistence contamination is expected to leave an imprint not only at small angular scales but also at the larger scales corresponding to the size of the field of view and to the periodicity of the observing pattern. This behaviour was already hinted at in the small-scale analysis, where the angular metric showed significant deviations from zero even at the largest angular separations beyond the main peak.

This effect becomes clearer when considering the galaxy--galaxy angular correlation at large scales. Since persistence sources are repeated following the sequence of telescope pointings, they introduce an angular periodicity that propagates into the measured galaxy auto-correlation, with an amplitude that depends on the contamination fraction. This behaviour is illustrated in Fig.~\ref{fig:ls-angstat-f10f50}-\textit{top}, where the light blue curve corresponds to $f=10\%$ and the orange one to $f=50\%$. The error bars represent the scatter among the mock realizations. Dashed vertical lines represent the first $N$ periodicities of the field of view angular scale\footnote{This value is slightly smaller than the nominal \ang{0.7} due to the overlaps between the dithers of consecutive groups of pointings.}. In the case of $f=10\%$, the oscillatory pattern is strongly diluted, while for $f=50\%$ it becomes clearly visible. In addition to these oscillations, the smooth component of the galaxy auto-correlation is also modified. Indeed, the measured galaxy--galaxy correlation is the combination of two contributions: the true clustering of target galaxies and the auto-correlation of persistence sources (shown in Fig.~\ref{fig:persistence-overview} in Appendix~\ref{app:nonmeascorrelations-persistence}), weighted by the squared fractions of the two populations in the sample. For very large contamination fractions, this additional component can dominate the signal and strongly suppress cosmological clustering signatures like the BAO peak. 
%at $2.4\degree$ in $z_1$ for \textbf{this} reference cosmology)

The star--galaxy correlation is also enhanced in the presence of strong persistence contamination. However, the oscillatory pattern visible in the persistence--persistence and galaxy--galaxy correlations is not clearly visible in this statistic. This happens because the star--galaxy correlation compares persistence signals, which are displaced according to the dithering pattern, with the original sky positions of the stars, rather than correlating persistence signals that share the same relative offsets.

Both the galaxy--galaxy and star--galaxy angular correlations for $f=50\%$ exhibit a zero-crossing around $10\degree$. For $f=10\%$, the galaxy auto-correlation crosses zero at approximately $2\degree$, while the star--galaxy correlation again crosses zero near $10\degree$. The same star--star correlation becomes negative at approximately $10\degree$. This behaviour contributes to the overall distortion observed in the contaminated angular statistics, while the oscillatory pattern in the galaxy--galaxy correlation originates from the repetition of persistence signals associated with the sequence of \Euclid\ pointings. 
%\textcolor{magenta}{0-crossing at $10\degree$: real effect, or due to the finite dimension of the S1 field?}

%%%%%%%%%%%%%%%%%%%%%%%%%%%%%%%%%%%%%%%%%%%%%%%%%%%

\section{\label{sc:starmask}Modelling and measuring star-mask inconsistencies}

\subsection{Star-mask mismatches between data and random catalogue}

The \Euclid MER pipeline applies masks around bright stars to remove spurious photometric detections that originate from diffraction spikes and other artifacts in the imaging \citep{Q1-TP004}. This produces holes in the survey coverage that must be properly accounted for in the random catalogue for galaxy clustering. The detection efficiency in photometry and spectroscopy is also reduced near to bright stars due to the point spread function and higher background level, an effect which is traced in the data but not directly in the random catalogue. To account for these effects, the masking of the data and random catalogues must be well-calibrated.

In this work, we assessed the systematic error induced in the galaxy clustering statistics when there is not a perfect match between the effective masking around bright stars in the data and random catalogues.
In particular, we modelled the stellar mask as a circular region centred on each star, but we allowed the mask radius to differ between the data and the random catalogue. This mimics the effects in real data, namely the reduced efficiency of detecting a galaxy near to a bright star, and any geometric difference in the mask applied to the data and random catalogue.

\subsection{MER masking procedure}
\label{sec:mermasking}

\begin{comment}
\begin{figure}
    \centering
    \includegraphics[width=\linewidth]{Figures/Mask/mag_vs_radius_mer.png}
    \caption{OU-MER magnitude vs. cut-out radius relation, in different units of angular separations. \textcolor{gray}{Probably not necessary.}}
    \label{fig:mag-vs-rad}
\end{figure} 
\end{comment}

We replicated the bright star mask constructed by MER on mock galaxy catalogues. In this work we neglect the mask applied around diffraction spikes, because the area is much smaller than that associated with the central circular mask.

% OU-MER masks bright stars using a set of polygons centred on the stellar position. In practice, this masking consists of a circular region whose size depends on the star's \Gaia\ magnitude, combined with a geometrical model that accounts for the diffraction spikes. As shown in \citet{Q1-TP004}, the area lost due to the diffraction spikes is much smaller than that associated with the central circular mask. For this reason, and in order to implement a fast and flexible approximation of the MER masking strategy in our catalogues and mocks, we neglect the contribution from diffraction spikes. 

We report here the criterion adopted by OU-MER to define the radius of the circular mask applied around a star as a function of its magnitude:
\begin{eqnarray}
\label{eq:MER-mask-mag-vs-radius}
    m_{\rm ref} \in& \{8.0, 9.0, 10.0, 11.0, 13.0, 15.0, 17.0, 20.0\} \, ,\nonumber\\
    \theta_{\rm ref} \in& \{400, 250, 150, 130, 60, 35, 17, 10\} \, , \nonumber
\end{eqnarray}
where the magnitudes are \Gaia\ $G_{\mathrm{RP}}$ magnitudes and the unit of the circular radius is the number of MER pixels, each one corresponding to $\ang{;;0.1}$. As our baseline we use a masking radius of 2 times the MER reference radius listed above (hereafter `MERCircx2') because this was found to better match the halos around bright stars in real data.
%Figure \ref{fig:mag-vs-rad} graphically shows this relation.

The masking procedure for each star is done as follows:
\begin{itemize}
    \item we interpolate the magnitude versus radius relation in correspondence of the magnitude of the stars. If the magnitude falls beyond the limits, we extend the highest masking only to the brightest stars, whereas we do not mask at all sources fainter than $m=20$;
    %\item we convert the angular size of the mask into Cartesian units on the unitary sphere; \textcolor{magenta}{(probably too specific)}
    \item we mask galaxies around each star within its corresponding masking radius.
\end{itemize}

\subsection{Setting up the angular correlation analysis for star mask}

\begin{figure}
    \centering
    \includegraphics[width=.9\linewidth]{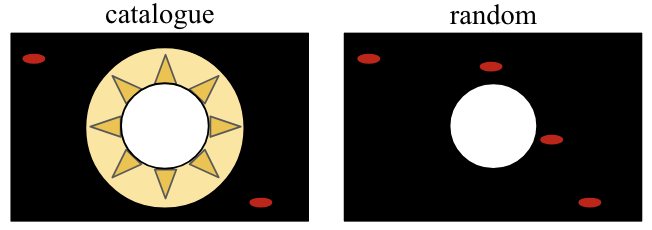}
    \caption{Simplified doodle of the mismatch between the star mask applied to the data catalogue (\textit{left}) and to the random catalogue (\textit{right}). If the MER star mask (white circle) is underestimated, there will be a noisy area around the mask which in practice will prevent sources to appear in the catalogue after applying a selection to the data. This information is not propagated to the random though, which results under-masked with respect to the data catalogue.}
    \label{fig:doodle-double-mask}
\end{figure}

To quantify the effect of a mismatch in the stellar masking applied to the data and random catalogues, we measured its impact on the star--galaxy angular correlation function by applying circular masks with different radii around stars in the two samples.  
We tested several masking mismatches, including the case in which the random catalogue was more strongly masked than the data around stars. As a reference configuration, we considered the case in which neither the data nor the random catalogue was masked. We then compared it with the MERCircx2 configuration, that is a masking applied following the MER strategy but with a radius twice as large. Finally, we introduced different masking radii in the random catalogue relative to the MERCircx2 mask applied to the data, up to the limiting case in which no masking was applied to the random catalogue. 

The star catalogue was constructed by selecting sources from the original \Gaia\ catalogue containing the S1 field and restricting it to the S1 footprint. 
%This was obtained by matching the \Gaia\ catalogue to the union of the random catalogues in all four spectroscopic redshift bins through an healpix map of $NSIDE=4096$. \textcolor{gray}{See persistence section: here the resolution is double, since I was using the numbers used by Pigi to build the footprint for the mocks. However, I checked that the difference in the number of stars with the two resolution is $2\%$.} 
As random catalogue for the stars, we used the random catalogue of the target galaxies without applying stellar holes.
We used {\tt corrfunc} to compute the star--galaxy correlation functions, evaluating them in 30 logarithmically spaced bins spanning the range $\theta \in [1\arcsec, 1\degree]$.

\subsection{Results on star mask}

\begin{figure}
    \centering
    \includegraphics[width=\linewidth]{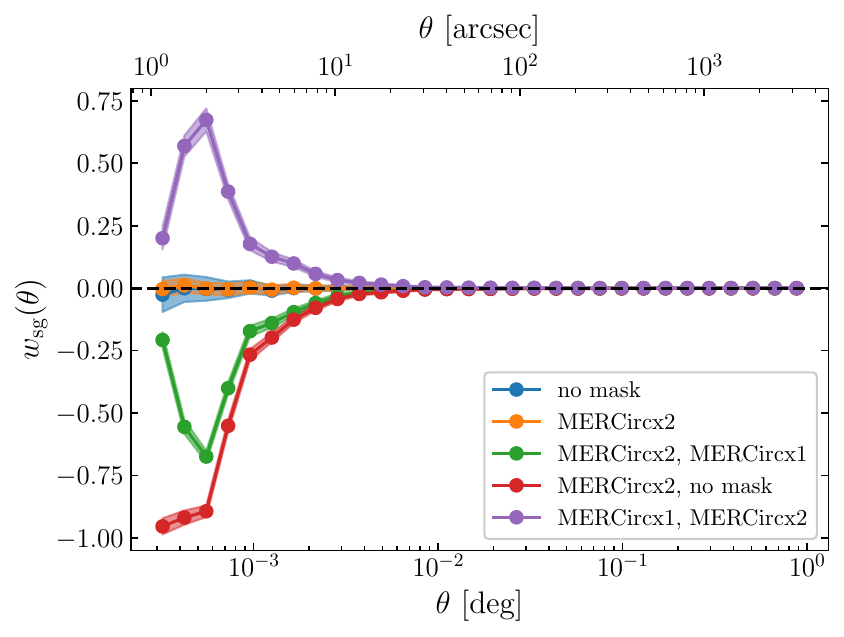}
    \caption{Star--galaxy angular correlation in different star mask scenarios, averaged over the 100 ELM. The error bars are the scatters over the mock measurements. The blue and orange line correspond to cases in which the same mask was applied to the data and to the random catalogue; in the purple case, the holes around stars in the random are bigger then in the data; in the green and red case, the random is less masked compared to the data, up to the limit case (red curve) where the random is not masked at all.}
    \label{fig:stargal-mask}
\end{figure}

\begin{figure}
    \centering
    \includegraphics[width=\linewidth]{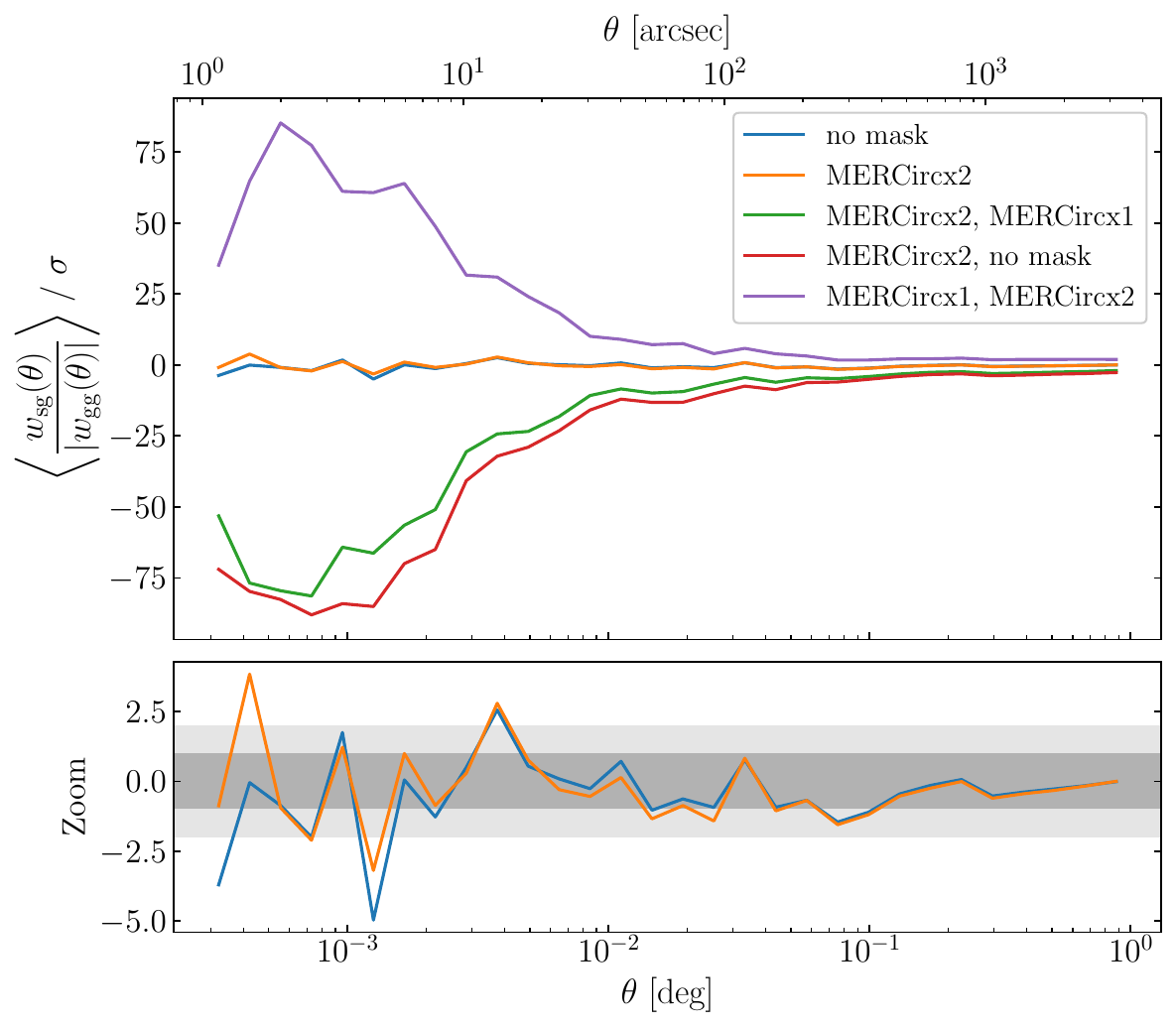} %\hfill
    \caption{Angular metric for each star mask configuration (equivalent to Fig.~\ref{fig:persistence-metric-significance}).}
    \label{fig:mask-metric-significance}
\end{figure}

Figure \ref{fig:stargal-mask} shows the star--galaxy cross-correlation measured for the different masking configurations considered in this test. The errors correspond to the scatter over the 100 ELM measurements. %\textcolor{gray}{Should we discuss why the no-mask case is not exactly 0 at small scales? It appears in the metric as well: see the same discussion on persistence, it may be due to star--star.}. 
The blue line corresponds to the case in which neither the data nor the random catalogue is masked. As expected, no cross-correlation between stars and the mock galaxies is detected. As expected, the same result (no cross-correlation) is obtained when the same mask is applied consistently to both the data and the random catalogues (orange curve). We then considered three possible cases of mismatched masking. The purple line corresponds to the case in which the holes in the random catalogue are larger than those in the data by a factor of two, with the random catalogue masked using a radius twice as large as that applied by MER to the data. The green line represents the opposite situation, where the random catalogue is less strongly masked than the data. Finally, the red line shows the extreme scenario in which the data are masked using the MERCircx2 configuration, while no masking at all is applied to the random catalogue.

A clear departure from zero appears in the star-galaxy cross-correlation when the masking applied to the data and to the random catalogue is not consistent. 
The characteristic shape of the cross-correlation curves can be understood in terms of the geometry of the stellar mask in the different configurations. 

As an illustrative example, let us consider the configuration shown in Fig.~\ref{fig:doodle-double-mask}, where the random catalogue is under-masked with respect to the data, corresponding to the green curve in Fig.~\ref{fig:stargal-mask}. In this case, three radial regimes can be identified around each star. In the innermost region ($\theta < \theta_1$), both the data and the random catalogue are masked, so the behaviour is the same as in the case where the two catalogues are masked consistently (blue and orange curves). In the intermediate annulus ($\theta_1 < \theta < \theta_2$), random galaxies are present while data are not, producing the same qualitative behaviour as in the limiting case where the random catalogue is not masked at all (red curve). Finally, beyond the outer boundary ($\theta > \theta_2$), neither catalogue is masked and the cross-correlation is expected to return to zero.
This behaviour can be interpreted more quantitatively using the natural estimator for the angular cross-correlation \citet{Peebles&Hausser1974}
\begin{align}
w(\theta) \approx \dfrac{\mathrm{D}_{\mathrm{g}} \mathrm{D}_{\mathrm{s}} (\theta)}{\mathrm{R}_{\mathrm{g}} \mathrm{R}_{\mathrm{s}}(\theta)} - 1 \, ,
\end{align}
to which the LS estimator reduces in the limit of infinite random sampling. In the central masked region, both $\mathrm{D}_{\mathrm{g}} \mathrm{D}_{\mathrm{s}}$ and $\mathrm{R}_{\mathrm{g}} \mathrm{R}_{\mathrm{s}}$ vanish apart from those of the faintest unmasked stars, which are consistently (non) masked. In the intermediate annulus, where $\mathrm{D}_{\mathrm{g}} \mathrm{D}_{\mathrm{s}} = 0$ but $\mathrm{R}_{\mathrm{g}} \mathrm{R}_{\mathrm{s}} \neq 0$, the estimator tends to $-1$, corresponding to a strong anti-correlation. Outside the masked region, where both data and random pairs are again present, the excess of random pairs disappears and the cross-correlation tends back to zero.
%Any deviation of the star--galaxy cross-correlation from $-1$ in the first bin $[1\arcsec, 1.3\arcsec]$ for the unmasked random case is due to the unmasked stars, i.e stars with $m>20$, which are not masked by OU-MER but still contribute to the star--galaxy cross-correlation signal. 
The measured correlations appear smoother than this idealized picture because stars with different magnitudes are masked with different radii, which broadens the sharp transitions between the different regimes. The behaviour of the star--galaxy correlation in the other masking configurations can be interpreted in an analogous way.

Figure \ref{fig:mask-metric-significance} shows the angular metric corresponding to each masking configuration. We can see that the significance of the metric in the presence of a mismatch in the stellar mask is very high. When rescaled to DR1-like uncertainties, the deviation from zero corresponding to the largest mismatch (at angular scales around $4\arcsec$) exceeds $20\sigma$. This indicates that such an effect should be clearly detectable in the Euclid DR1 data.
%\textcolor{magenta}{(We did it in RR2 and TR1! Should we anticipate this here as part of a DR1 paper in prep.?).}

\section{\label{sc:largescales}Effect of star systematics on 2PCF}

\begin{figure}
    \centering
    \includegraphics[width=\linewidth]{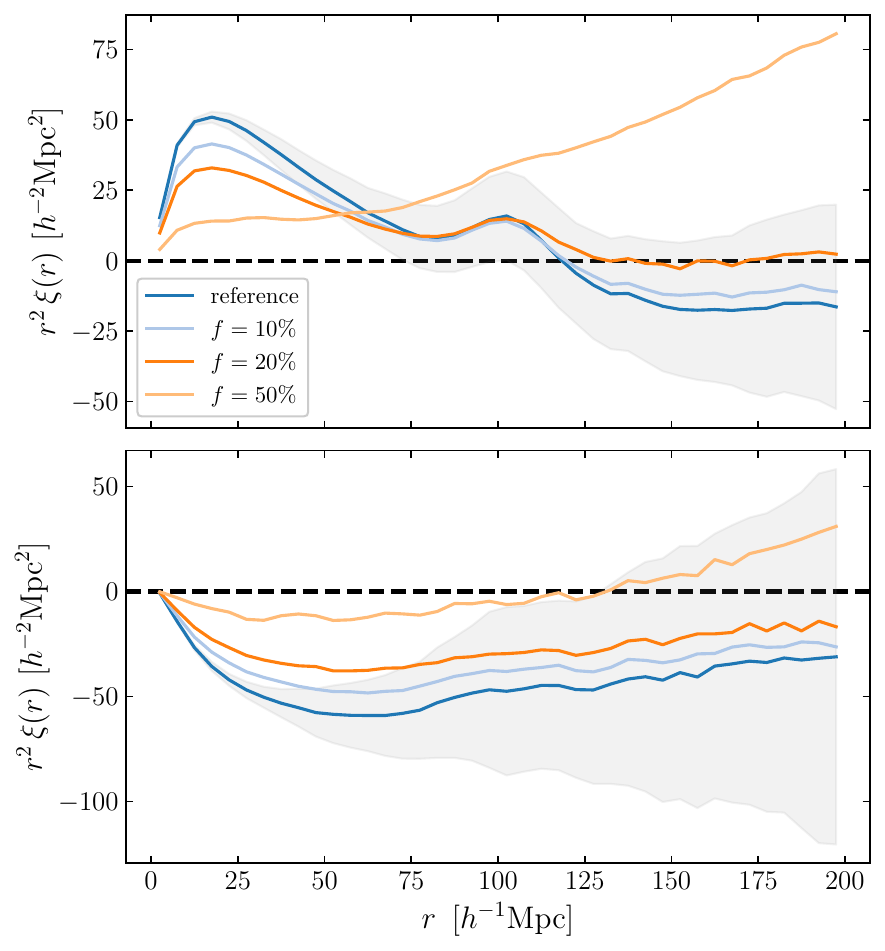}
    \caption{Monopole (\textit{top}) and quadrupole (\textit{bottom}) of the 2PCF of the target galaxies (blue line) and of the contaminated catalogues in presence of different fractions of persistence. The shaded blue area represents the uncertainty on the single measurement for the target reference case on S1.}
    \label{fig:ls-2pcf-persistence}
\end{figure}

\begin{figure}
    \centering
    \includegraphics[width=\linewidth]{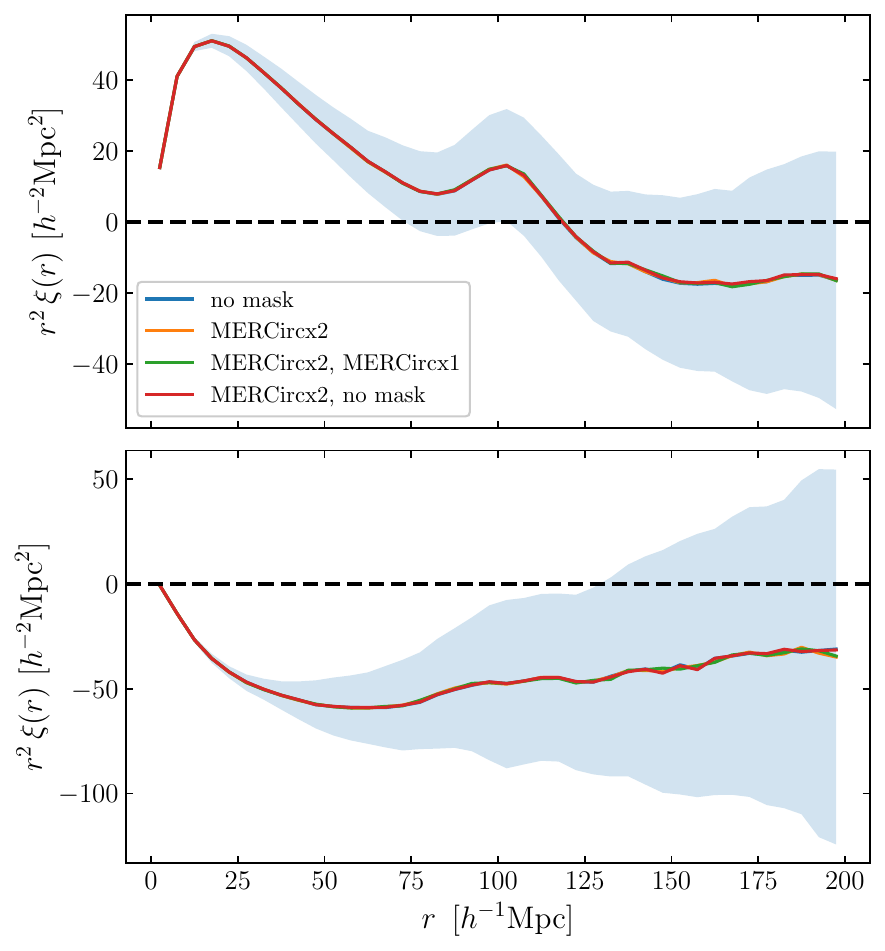}
    \caption{Monopole (\textit{top}) and quadrupole (\textit{bottom}) of the 2PCF of the coherent masking case case (blue line) and of the the other data and random masking configurations. The shaded blue area represents the uncertainty on the single measurement for the reference case on S1.}
    \label{fig:ls-2pcf-starmask}
\end{figure}

We have thus far focused on the angular correlations, given their sensitivity to star-induced systematics like persistence and masking mismatches. However, clustering analyses are fundamentally based on spatial correlation functions. It is therefore necessary to quantify the impact of star-related systematics on these statistics. In this work, we consider the three dimensional galaxy–galaxy two-point correlation function (2PCF).

As discussed above, persistence follows a well-defined and regular angular pattern set by the \Euclid\ pointing sequence. This gives rise to a population of persistence-driven contaminants with a characteristic autocorrelation function that differs, in both amplitude and shape, from that of the genuine galaxy population (see Appendix \ref{app:nonmeascorrelations-persistence}). Persistence-induced contamination is therefore expected to modify the angular clustering properties of the sample, while only diluting the radial clustering due to the random redshifts assigned to contaminants.
%leaving the radial clustering largely unaffected.
Consequently, the spatial 2PCF of a contaminated sample will differ from that of an uncontaminated one.
We expect the main features of the contaminated spatial 2PCF to qualitatively mirror those observed in the angular case, including the presence of a zero-crossing at a physical scale corresponding to the angular scale at the relevant redshift.

In Fig.~\ref{fig:ls-2pcf-persistence}, we show the monopole and quadrupole of the 2PCF computed up to a pair separation of $200\,\hMpc$, for samples characterized by different levels of persistence contamination and for the uncontaminated case (blue curve). The primary effect of persistence, for contamination fractions $f \lesssim 20\%$, is a suppression of the clustering amplitude, as evidenced by the damping of the monopole (top panel), and a reduction of the anisotropy induced by redshift-space distortions, with the quadrupole approaching zero (bottom panel). The magnitude of these effects increases with the contamination fraction.
However, unlike the case of contamination by unclustered  noise interlopers (see \citealp[]{EP-Risso}), persistence-induced contamination also alters the shape of the monopole, not only its amplitude. This distortion is already apparent at $f = 20\%$, where a significant broadening of the BAO peak and a shift in the zero-crossing scale are observed. It becomes dominant in the extreme and likely unrealistic case of $f = 50\%$, where the amplitude of the correlation signal multiplied by  $r^2$, being entirely driven by persistence, increases monotonically  up to separations of $\sim 300\,\hMpc$. 
Beyond this separation, the monopole amplitude starts decreasing approaching zero around $450\,\hMpc$. This scale approximately corresponds, at $z \sim 1$, to an angular separation of $\sim 10^\circ$, which is the zero-crossing scale of the contaminated angular two-point function.
It is therefore evident that the presence of persistence induced interlopers represents a potential risk for the galaxy clustering analyses and that keeping the contamination level below 10\% is of paramount importance.

By contrast, we found that the impact of an imperfect masking of the stellar population should be negligible for the DR1 spatial clustering analysis. In none of the cases explored, including the extreme scenario in which stellar masking is not accounted for at all in the random catalogue, the multipole moments of the 2PCF are significantly affected, as shown in Fig.~\ref{fig:ls-2pcf-starmask}.
This is not unexpected, as imperfect star masking is a local effect that primarily impacts sub-arcminute angular scales (see Fig.~\ref{fig:stargal-mask}), corresponding to sub-Mpc separations at the mean redshift of the sample ($z \sim 1$). We verified that the systematic errors induced on 2PCF by an imperfect masking contribute negligibly to the total error budget, remaining below the 10\% of the statistical uncertainties expected for the DR1 sample.
We observed a significant mismatch only in mock catalogues crossing the Galactic plane, that is, in regions characterized by extremely high stellar densities. 
In that case, the cumulative masked-area mismatch becomes a significant fraction of the total survey area and an integral-constraint–like effect arise, modifying the overall amplitude of the 2PCF.
Since the EWS is designed to avoid the Galactic plane and to target regions at relatively high Galactic latitude $|b| \gtrsim 20\degree$, such extreme conditions are not expected to occur.

\section{\label{sc:Conclusion}Conclusions}

In this work, we have investigated the impact of two types of star-related systematic effects that can potentially affect clustering analyses of the \Euclid\ spectroscopic galaxy catalogue. The first effect is related to the persistence of the photometric imprint of bright stars in the spectroscopic exposures. The second arises from mismatches in the effective star mask in the galaxy catalogue and in the survey selection function, which we model through a catalogue of synthetic random objects.

All results presented here are obtained by combining simulated galaxy catalogues with real stellar catalogues from the \Gaia\ and \twomass\ surveys. To characterise the impact of these effects on the upcoming Euclid DR1, we consider two-point statistics estimated from both the angular and spatial distributions of galaxies and stars.

In Sect.~\ref{sc:persistence}, we have shown that photometric persistence in spectroscopic images can represent an insidious source of systematic error for galaxy clustering analyses, while remaining readily identifiable with appropriate statistical tools. The persistence signal is characterized by a regular angular pattern and therefore exhibits a distinctive autocorrelation function, peaking at angular scales comparable to the dithering scale and extending to much larger separations.

Owing to its intrinsic correlation with the angular positions of the stars that generate the signal, persistence can be effectively detected through measurements of the angular star–galaxy cross-correlation function, which is expected to be consistent with zero in the absence of contamination. Our results show that the cross-correlation between a mock galaxy sample contaminated by persistence-induced interlopers and a stellar catalogue significantly deviates from zero, displaying two characteristic features shown in Fig.~\ref{fig:ideal-correlations}.
First, a pronounced bump is observed at $\sim 100\arcsec$, which can be regarded as a distinctive signature of persistence contamination. The significance of this feature, quantified using a dedicated metric, increases with the contamination fraction and, for a DR1-like sample, would be detectable even for contamination levels as low as $\sim 10\%$ (Fig.~\ref{fig:persistence-metric-significance}). Second, persistence induces a positive star–galaxy cross-correlation, whose amplitude correlates with the contamination fraction. This spurious excess is particularly evident on sub-arcminute scales and it remains potentially detectable out to separations of several degrees (Fig.~\ref{fig:ls-angstat-f10f50}).

From these results, the star–galaxy angular cross-correlation function clearly emerges as a powerful diagnostic for detecting a population of spurious objects misclassified as line-emitting galaxies due to incomplete removal of the persistence signal. This signature should not be overlooked, as our extension of the analysis to spatial clustering demonstrates that persistence-driven contaminants can significantly alter both the amplitude and the shape of the spatial 2PCF, potentially distorting the BAO feature depending on the contamination fraction (Fig.~\ref{fig:ls-2pcf-persistence}).

In Sect.~\ref{sc:starmask}, we have shown that inconsistencies in the stellar masking procedure can generate a spurious star–galaxy cross-correlation signal, which may be either positive or negative depending on the relative extent of the masked regions in the galaxy and corresponding random catalogues (Fig.~\ref{fig:stargal-mask}). In this case, however, the effect is highly localized, being confined to angular scales of a few tens of arcseconds around stars.
For this reason, and assuming to be sufficiently far from the Galactic plane as the \Euclid EWS is designed to be, the star-mask mismatch does not significantly impact the 2PCF measurements, even in the extreme scenario where stellar masking is not included at all in the random catalogues (Fig.~\ref{fig:ls-2pcf-starmask}).

Our results provide a clear indication of the potential impact of systematic effects related to the presence of stars in the \Euclid\ survey area. However, our predictions for the magnitude of these effects rely on simulated data and a number of simplifying assumptions.
Most importantly, in modelling persistence decay, we adopted a simplified phenomenological description of the signal evolution compared to Euclid Collaboration: Kubik et al. (in prep.).
%\citet{DR1-TP038}. 
We neglected the presence of multiple decay regimes in the power-law decay, and we adopted an effective description in which the persistence timescale, and hence the number of persistence images, is independent of both the magnitude of the stellar source and the position of the pixel on the focal plane. 
We also ignored the different modelling required for saturated sources and their long-term impact on the temporal evolution of the persistence signal, and we approximated the detector response using power-law coefficients calibrated on a single reference incident flux $I_0=30 \, \mathrm{k}e^-$. 
In addition, we neglected the intrinsic clustering of the photometric galaxies when adding the persistence contaminants to the spectroscopic galaxy sample. 
While these assumptions provide a conservative and computationally tractable framework for an initial assessment, a fully realistic treatment will require end-to-end simulations and a detailed calibration of all relevant effects using early survey data. These aspects, along with a quantitative assessment of the impact of residual persistence on cosmological parameter estimation, will be addressed in future work.              
%(I'd conclude the paper here)

In summary, angular two-point statistics provide a robust and efficient framework to detect and characterize star-related systematics in \Euclid spectroscopic data. Stellar mask mismatches, although easily detectable at small scales, are cosmologically harmless in realistic configurations. Photometric persistence, on the other hand, can induce scale-dependent distortions in clustering measurements if left uncorrected. Crucially, the small-scale star–galaxy angular cross-correlation acts as a direct and sensitive probe of such contamination: if no significant deviation from zero is observed, large-scale clustering analyses, including BAO measurements, can be considered robust against star-related systematic effects. 

Dedicated papers will present the analysis of observational systematics in the DR1 data. 
%In particular, the Training Release 1 (TR1), corresponding to an unblinded subset of DR1 data, comprises two separate regions in the Northern and Southern hemispheres, covering a total area of approximately $500,\mathrm{deg}^2$. TR1 was specifically designed within the Euclid Collaboration to characterize observational systematics and calibrate the catalogue selection function, providing a representative dataset on which these effects can be investigated without accessing the blinded DR1 sample. Preliminary measurements on TR1 confirm the presence of star-mask mismatches. In some redshift bins, we also observe a mild excess of power around the characteristic persistence scale, which can be mitigated through an appropriate colour selection of the spectroscopic catalogue. In addition, we identified a different star-related systematic associated with spurious source detections around stellar halos and diffraction spikes. Although this effect was not explicitly modelled in this work, its angular signature resembles that produced by an under-masked random catalogue, illustrating how the diagnostics developed here can also help identify systematic effects beyond those explicitly considered in our simulations.}
More generally, spurious clustering features on scales comparable to the \Euclid field of view can be generated by observational effects other than persistence. The random catalogue, incorporating the VMSP survey selection function, will account for a number of these angular effects (Euclid Collaboration: Granett et al. 2027, in prep.), while the diagnostics presented in this work provide a complementary means of assessing residual systematics and detecting effects, such as persistence, not modelled in the random catalogue. Extending this approach from configuration to harmonic space may provide complementary information and additional leverage for distinguishing between different sources of contamination. Work along these lines is currently in progress.

%
% Add the acknowledgement using the achnowledgements environment.
% Do not use \acknowledgement{....} as this affects the formatting
% of the references.
%

\begin{acknowledgements}
%\AckERO  
\AckEC  
\AckCosmoHub\
Simulations and computations in this work have been run at the computing facilities of INFN, Sezione di Genova: the authors wish to thank the INFN IT personnel in Genova for their precious and constant support. The authors acknowledges support from MIUR, PRIN 2022 (grant 2022NY2ZRS 001). 
%This work has made use of CosmoHub, developed by PIC (maintained by IFAE and CIEMAT) in collaboration with ICE-CSIC. It received funding from the Spanish government (grant EQC2021-007479-P funded by MCIN/AEI/10.13039/501100011033), the EU NextGeneration/PRTR (PRTR-C17.I1), and the Generalitat de Catalunya \citep{FS2_2017ehep.confE.488C, FS2_TALLADA2020100391}.
This work has made use of data from the European Space Agency (ESA) mission Gaia (\url{https://www.cosmos.esa.int/gaia}), processed by the Gaia Data Processing and Analysis Consortium (DPAC, \url{https://www.cosmos.esa.int/web/gaia/dpac/consortium}). Funding for the DPAC has been provided by national institutions, in particular the institutions participating in the Gaia Multilateral Agreement.
This publication makes use of data products from the Two Micron All Sky Survey, which is a joint project of the University of Massachusetts and the Infrared Processing and Analysis Center/California Institute of Technology, funded by the National Aeronautics and Space Administration and the National Science Foundation.
“ELSA: Euclid Legacy Science Advanced analysis tools” (Grant Agreement no. 101135203) is funded by the European Union. Views and opinions expressed are however those of the author(s) only and do not necessarily reflect those of the European Union or Innovate UK. Neither the European Union nor the granting authority can be held responsible for them. UK participation is funded through the UK Horizon guarantee scheme under Innovate UK grant 10093177.
\end{acknowledgements}

%
% Here comes the reference list, generated via bibtex from
% your bibfile my.bib and Euclid.bib. Please make sure that
% the same paper is not referenced twice, one from your my.bib
% file, and once from Euclid.bib.
%

\bibliography{Euclid, mybib, DR1} % add my.bib, containing your bibentry file 

%
% Now you can add appendices.
% In this example, the appendices are in one column mode.
% If that is not requires, comment out \onecolumn
% Note that appendices in A\&A come {\it after\/} the references.

%%%%%%%%%%%%%%%%%%%%%%%%%%%%%%%%%%%%%%%%%%%%%%%%%%%%
%%%%%%%%%%%%%%%%%%%%%%%%%%%%%%%%%%%%%%%%%%%%%%%%%%%%

\begin{appendix}
  %\onecolumn %If you don't want single column for the Appendix, please
             %comment this out

\section{\Euclid ROS scheme}
\label{app:ROS}

\begin{figure*}
    \centering
    \includegraphics[width=0.9\linewidth]{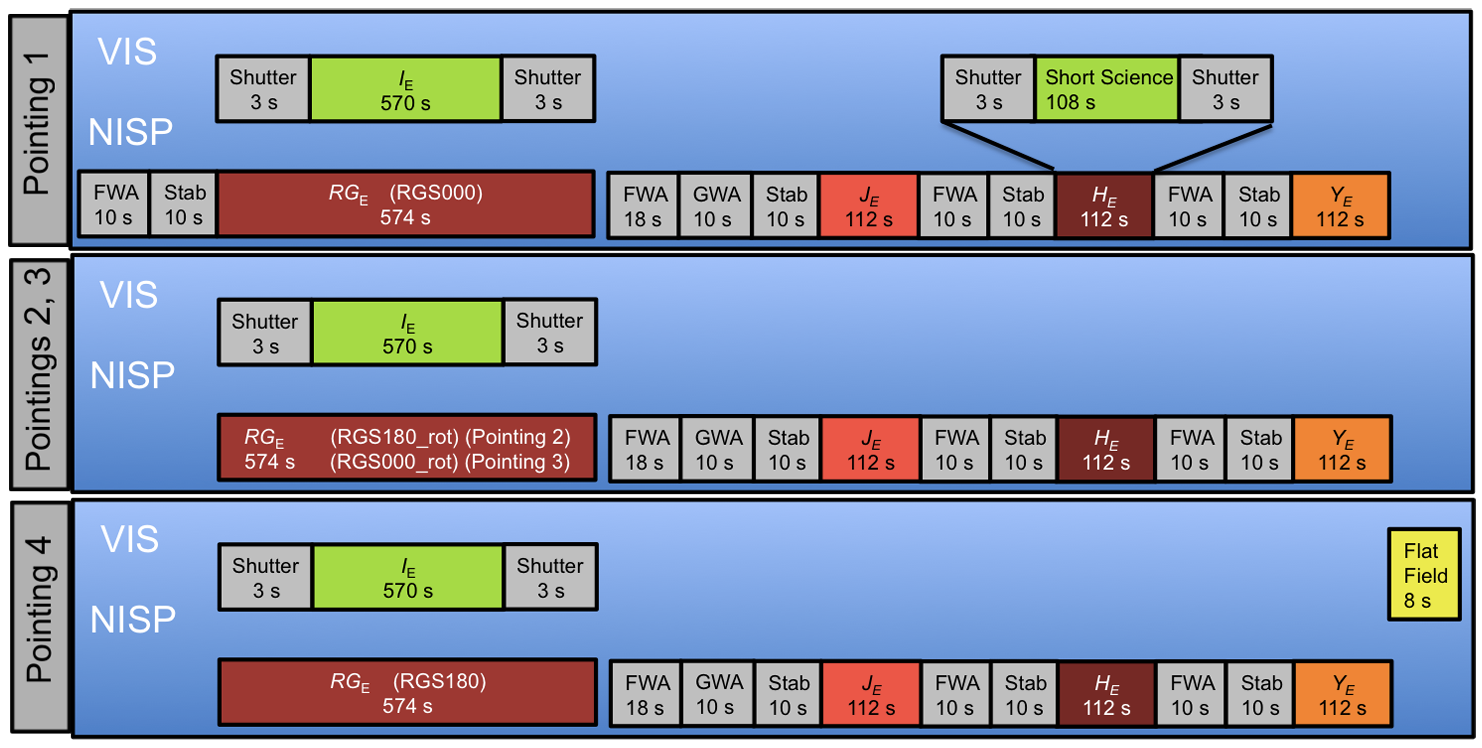} 
    \caption{Scheme of the \Euclid ROS, as reported in figure 8 of \cite{Scaramella-EP1}.}
    \label{fig:surveysequence}
\end{figure*}

We include a picture of the \Euclid observation sequence in Fig.~\ref{fig:surveysequence} for the reader's convenience. In each pointing (or dither), a sequence of spectroscopic and photometric exposures is performed using the NISP instrument. In each dither, the spectroscopic exposure is performed using a specific grism with a given offset angle: this sequence is repeated in the same order in all groups of 4 dithers, as reported in \cite{Q1-TP006}. All dithers share the same sequence of photometric exposures performed in the three \Euclid near-infrared bands.

\section{From \twomass\ magnitudes to \Euclid electron counts}
\label{app:from-mag-to-ne}

In order to apply the persistence model described in Sect.~\ref{sc:persistence-model} at the detector level, we must convert stellar magnitudes into the corresponding number of photo-electrons expected in the \Euclid photometric images. For this purpose, we relied on the PSC \twomass\ catalogue, whose near-infrared bands are closely matched to those of \Euclid. This choice allows for a straightforward transformation between \twomass\ magnitudes and the \Euclid photometric system, as detailed in \cite{Schirmer-EP18}.
Since \twomass\ magnitudes are measured in the Vega system,
%before translating them into the \Euclid bands, 
we first transformed them into the AB system using Table~$1$ of \cite{Blanton_2007},
\begin{align}
    J_{\mathrm{2MASS,AB}} &= J_{\mathrm{2MASS,Vega}} + 0.91 \, ,\\ 
    H_{\mathrm{2MASS,AB}} &= H_{\mathrm{2MASS,Vega}} + 1.39 \, . 
\end{align}
%and then convert these %\twomass\ 
%magnitudes into \Euclid bands using the equations reported in 
%Appendix E of \citealt{Schirmer-EP18}:
We then followed \cite{Schirmer-EP18} (Appendix E) to convert them into \Euclid near-infrared bands
\begin{align}
\JE &= H_{\mathrm{2MASS}} - 0.007 + 0.787\,(J_{\mathrm{2MASS}} - H_{\mathrm{2MASS}}) \, ,\\ 
\HE &= H_{\mathrm{2MASS}} + 0.069 - 0.119\,(J_{\mathrm{2MASS}} - H_{\mathrm{2MASS}}) \, ,\\ 
\YE &= H_{\mathrm{2MASS}} - 0.005 + 1.134\,(J_{\mathrm{2MASS}} - H_{\mathrm{2MASS}}) \, .
\end{align}
 
Here we considered only the last of the three photometric exposures within each dither, corresponding to the \YE band. In the absence of relative offsets between consecutive photometric exposures, the persistence signals generated in the \JE and \HE bands would coherently accumulate with that of the \YE exposure in the same detector pixels, nearly doubling the residual persistence signal in the subsequent spectroscopic frame. However, the typical jitter between consecutive photometric exposures exceeds the pixel scale, spatially decorrelating the persistence patterns. We therefore neglect the contribution from the \JE and \HE exposures.  

To compute the number of electrons (both absolute and per second) generated by a star of a given \Euclid magnitude $m$, we referred to the equations and ZP values reported in \cite{Schirmer-EP18},
\begin{align}
    %m_{\rm AB} &= -2.5 \, \log{ \left( \frac{f_{\nu} (\nu) }{1Jy} \right)} + 8.9 \\
    n_{\mathrm{e}} \,/\, t_{\mathrm{photo}} &= 10^{-0.4\, (m-\mathrm{ZP}_{\HE})} \,\mathrm{e}^-\,\mathrm{s}^{-1}\, ,\\
    \mathrm{ZP}_{\HE} &= 25.21  \, ,
\end{align}
considering $t_{\mathrm{photo}}=112 \,\mathrm{s}$ as duration of the photometric exposure, during which the star light is gathered on the FP and generates $n_{\mathrm{e}}$ electron counts.

\section{Modelling the stellar images and treatment of saturated stars}
\label{app:I0modelling}

%In order to apply Eq.~\eqref{eq:pers-current}, we must determine the value of $I_0$ in the pixels that contain the image of the star, which, in the \Euclid case, is described by a Gaussian point spread function (PSF) with $\mathrm{FWHM} = \ang{;;0.49}$ \citep{EROData}.
%Cuillandre+2025
%Estimating the persistence signal from Eq.~\eqref{eq:pers-current} requires specifying $I_0$ in all pixels containing the stellar image. This can be modelled by convolving the photon flux with a 2-dimensional Gaussian point spread function (PSF) with $\mathrm{FWHM} = \ang{;;0.49}$ \citep{EROData}. The stellar magnitude then determines both the number of pixels in the image and the fraction that is saturated.

In order to apply Eq.~\eqref{eq:pers-current}, we must determine the value of $I_0$ in the pixels that contain the image of the star. Since stellar images in \Euclid are described by a Gaussian point spread function (PSF) with $\mathrm{FWHM} = \ang{;;0.49}$ \citep{EROData}, we use the same model to distribute the stellar photon flux across detector pixels. The stellar magnitude then determines both the number of illuminated pixels and the fraction that becomes saturated.

%%% Too verbose, but detailed
%After centring the position of the star in one of the detectors' pixels, we computed the number of electrons $n_{\mathrm{e}}$ collected in the central pixel and in the surrounding ones.
%If the star is sufficiently bright, the signal in some pixels saturates; moreover, the brighter the star, the greater the number of saturated pixels. For example, a star with magnitude $m = 12$ saturates approximately $4 \times 4$ pixels, while a star with $m = 16$ saturates only the central pixel; fainter sources do not produce saturation.

However, for simplicity we neglected any difference in the extension of the saturated signal depending on the star magnitude. We only focused on the number of electrons falling in the central pixel of the star, and we compared it to the full well (FW) capacity of the NISP detectors, that is, the maximum number of electrons a pixel can store before saturation occurs ($130 \,\mathrm{ke}^-$ according to \citealt{EU-Kubik}). We therefore adopted the following prescription.
If $n_{\mathrm{e}} < \mathrm{FW}$, we set $I_0 = n_{\mathrm{e}}$ as the initial signal entering the persistence computation. 
Conversely, if $n_{\mathrm{e}} > \mathrm{FW}$, we assumed $I_0 = 3 \, \mathrm{FW}$, adopting a conservative approach in which strong over-saturation leads to enhanced persistence. 
%\textcolor{magenta}{(Bogna uses 2 in the issue n° 31436: since it's arbitrary, we were more conservative )}.
This is a simplified treatment, as saturated and non-saturated sources are expected to follow different persistence behaviours, whose modelling remains uncertain (Euclid Collaboration: Kubik et al., in prep.).
%\citep{DR1-TP038}.

\section{Tests on Flagship}
\label{sc:FS2}

\begin{figure}
    \centering
    \includegraphics[width=1.01\linewidth]{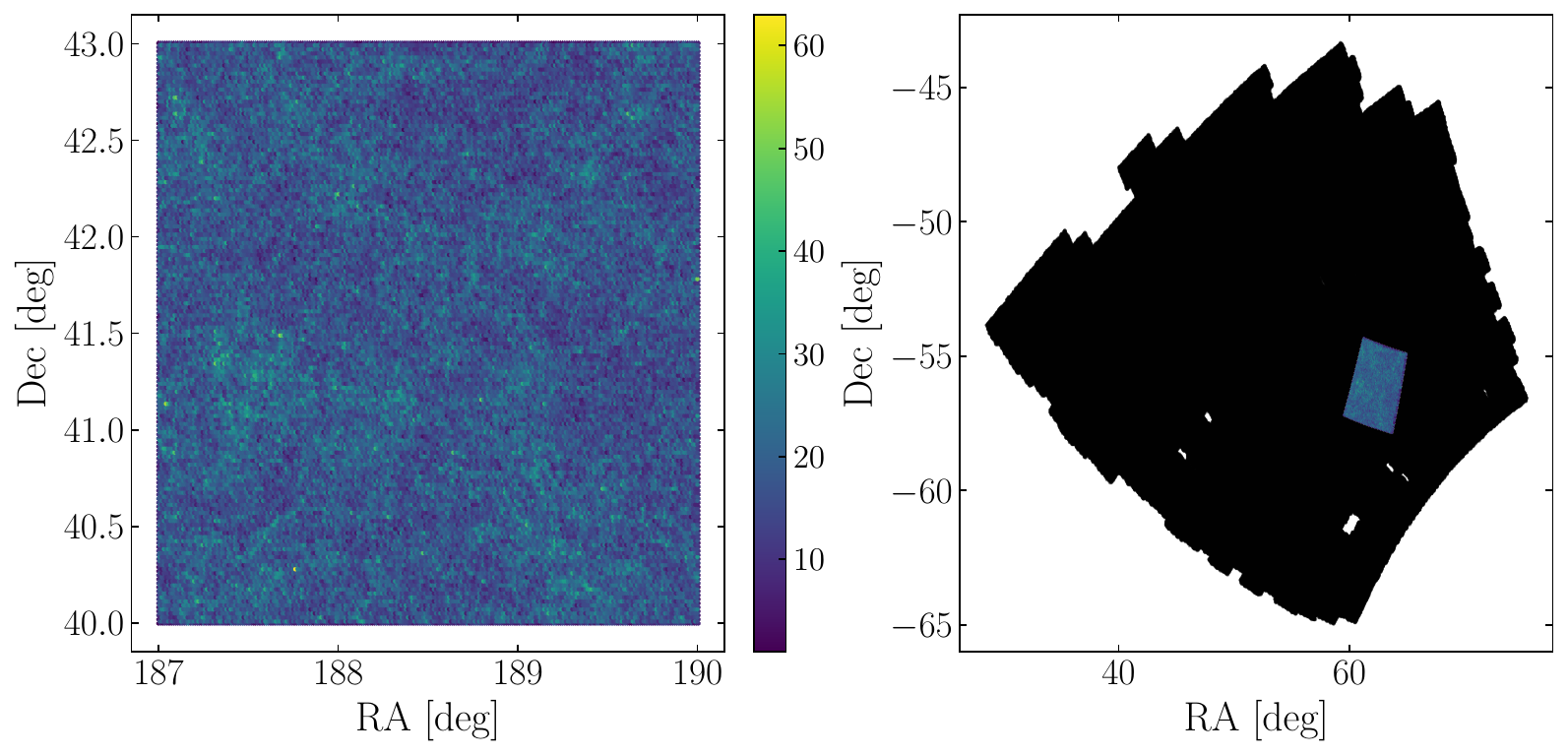}
    \caption{\textit{Left}: number density  of galaxies in the $3^{\circ} \times 3^{\circ}$ sub-patch selected from the Flagship simulation. The colour scale indicates the number of galaxies per hexagonal bin. \textit{Right}: location of the same sub-patch within the S1 footprint (shown in black).
    %The globular cluster in S1 is visible here as a yellowish spot around $\mathrm{RA}=49^{\circ}$ and $\mathrm{Dec}=55^{\circ}$.
    }
    \label{fig:FS23x3inS1}
\end{figure}

\begin{figure}
    \centering
    \includegraphics[width=\linewidth]{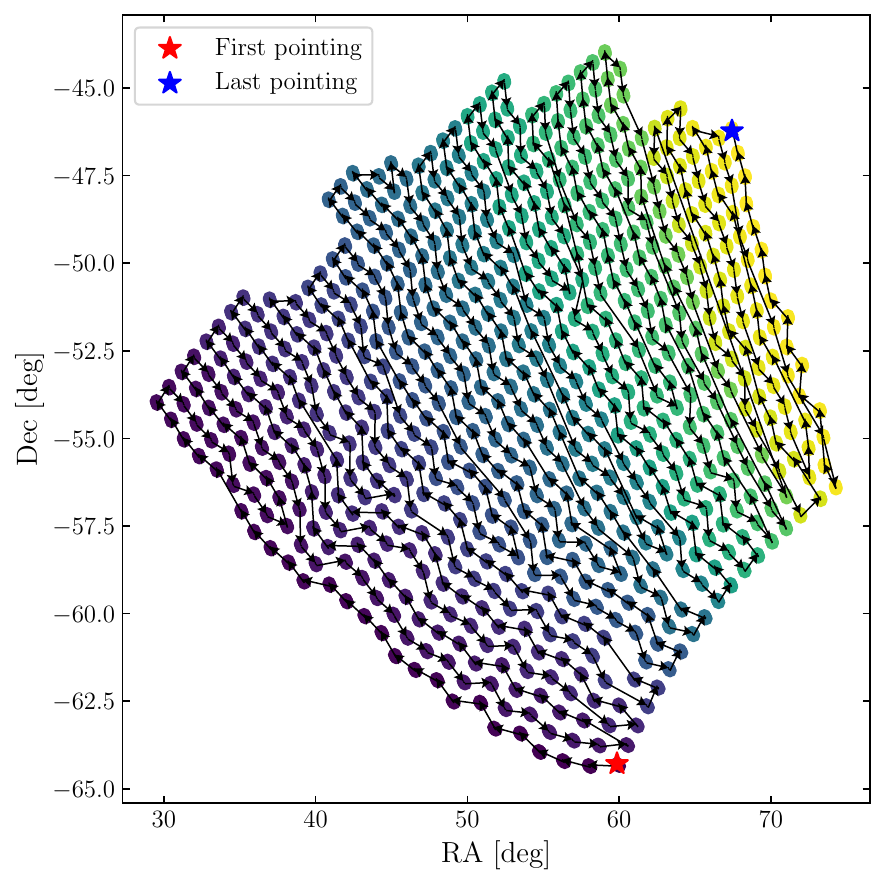}
    \caption{Centres of the pointings used to observe the S1 field. The arrows and the colour gradient indicate the progression in time of the observations.}
    \label{fig:S1-pointing-sequence}
\end{figure}

Before performing the persistence simulations on mock galaxy catalogues, we first assessed the expected frequency of overlaps between persistence signals and \Euclid\ galaxies, as well as the relative contribution of photometric galaxies and ELGs. In this context, we define ELGs as galaxies with $f_{\ha} > 2 \times 10^{-16}\,\mathrm{erg}\,\mathrm{cm}^{-2}\,\mathrm{s}^{-1}$. Since ELGs are expected to represent only about 2\% of the \Euclid\ photometric catalogue, we expect most overlaps to involve photometric galaxies.

For this test, we used the Flagship simulation \citep{EuclidSkyFlagship}, selecting galaxies with $\HE < 24$ as a proxy for the \Euclid\ photometric sample. To balance statistical robustness and computational feasibility, we restricted the analysis to a $3\degree \times 3\degree$ region ($187\degree < \mathrm{RA} < 190\degree$, $40\degree < \mathrm{Dec} < 43\degree$). This sub-patch was repositioned within the S1 footprint, avoiding the stellar overdensity associated with the globular cluster discussed in Appendix~\ref{app:globclusS1}, in order to ensure a representative stellar environment for the persistence simulations (see Fig.~\ref{fig:FS23x3inS1}).

We then used the {\tt GELSA} code to simulate the observation of this sub-patch through the S1 pointing sequence, accounting for persistence signals generated by \twomass\ stars during the following $\neff = 8$ spectroscopic exposures. For each S1 pointing, we constructed the corresponding photometric and spectroscopic {\tt GELSA} frames, which allow the conversion between sky coordinates and focal-plane (FP) coordinates. These frames were used to compute the FP positions of both stellar images, which generate persistence, and dispersed galaxy spectra, whose orientation depends on the grism and therefore on the dither configuration.

Persistence signals were generated by tracking the FP positions of stellar images over the preceding \neff\ photometric exposures. Although the analysis focuses on a $3\degree\times3\degree$ region, the full S1 pointing sequence must be considered, since the spatial distribution of persistence signals depends on the temporal ordering of the observations. Figure~\ref{fig:S1-pointing-sequence} shows the complete S1 observing sequence.
%In this case, we directly used the big rectangular 2mass catalogue containing the S1 patch, since we optimized the simulation in order not to lose time trying to observe stars which are too far away from the field of view.
%We built the list of photometric persistence signals of all the observed stars in the following exposures, keeping memory, for each observed frame,  of the positions of the star images on the focal plane in the previous 8 photometric exposures. Those signals are then summed in each current observed frame, and will be added on top of the galaxies spectra in the corresponding spectroscopic exposure.
 
We then simulated the observation of the galaxies in the selected sub-patch through all S1 pointings covering that region. For each galaxy, we computed the FP coordinates of its spectrum in every frame, sampling wavelengths between $1200\,\si{nm}$ and $1850\,\si{nm}$ in steps of $3\,\si{nm}$.
%and we computed the positions of their spectra on the FP for each S1 frame which covered that area of the sky. 
%Differently from the stars, we didn't need to simulate the observation of all galaxies in the S1 sky region, but we could consider only those in the small sub-patch. 

Finally, we combined the stellar persistence signals with the galaxy spectra in the corresponding spectroscopic exposures and evaluated their spatial overlap. An overlap was defined when the distance between a persistence image and any point sampling the galaxy spectrum was smaller than $3.6\,\si{nm}$, corresponding to a `3-pixel' criterion.
%The dense sampling of the wavelength positions in the spectra described above was necessary because, when evaluating the overlaps, we set a maximum distance of 3 pixels to define it. Therefore, the points used to sample the spectrum coordinates must be separated by a smaller distance. 

We found that 66\% of galaxies exhibit at least one overlap with a persistence signal. In principle, this could introduce a large number of photometric galaxies into the spectroscopic sample, potentially exceeding the number of genuine spectroscopic targets. In practice, the effect is mitigated by masking procedures and by the spectroscopic pipeline, which removes many contaminated objects during spectral extraction or through quality-flag selections. Nevertheless, an unknown fraction of residual persistence signals may survive these cuts, depending on the catalogue selection criteria.

Among the galaxies affected by persistence, 97\% are photometric galaxies, supporting the assumption that persistence contamination is dominated by overlaps with photometric sources. We verified that this result is stable across all redshift bins, indicating that the relative contribution of photometric and spectroscopic galaxies to persistence contamination is largely independent of redshift.

\section{Persistence-persistence and star--persistence at small scales}
\label{app:nonmeascorrelations-persistence}

\begin{comment}
\begin{figure}
    \centering
    \begin{subfigure}[t]{\linewidth}
        \centering
        \includegraphics[width=\linewidth]{Figures/Persistence/pers-pers.pdf}
        \caption{Persistence–persistence angular auto-correlation on sub-degree scales for $f=10\%$ and $f=50\%$. The shaded areas correspond to the scatters among the 100 mock measurements.}
        \label{fig:pers-pers}
    \end{subfigure}
    \hfill
    \begin{subfigure}[t]{\linewidth}
        \centering
        \includegraphics[width=\linewidth]{Figures/LargeScales/largescales-perspers-f10f50.pdf}
        \caption{Large-scale persistence–persistence angular auto-correlation for $f=10\%$ and $50\%$. The coloured bands are Poisson errors.}
        \label{fig:ls-perspers-f10f50}
    \hfill
    \begin{subfigure}[t]{\linewidth}
        \centering
        \includegraphics[width=\linewidth]{Figures/Persistence/modelling-stargal-f50.pdf}
        \caption{Star–galaxy angular correlation measured on mocks (orange curve) and its model for a contamination fraction $f = 50\%$ (grey dashed line). The shaded area corresponds to the scatter among the 100 mock measurements.}
        \label{fig:modelling-stargal-f50}
    \end{subfigure}
    \end{subfigure}

    \caption{Overview of persistence-related correlation measurements and models.}
    \label{fig:persistence-overview}
\end{figure}
\end{comment}

Having presented the results of the auto- and cross-correlations of stars and galaxies in persistence-contaminated samples in Sects.~\ref{results-persistence-ss} and \ref{results-persistence-ls}, we report here for completeness the angular auto-correlation of persistence contaminants $w_{\mathrm{pp}} (\theta)$. 
%and the cross-correlation between persistence contaminants and stars ($w_{\mathrm{pp}} (\theta)$). 
Although this quantity is not directly measurable in real data, it provides useful insights into the interpretation of the observable galaxy--galaxy and star--galaxy correlation functions in contaminated samples.

The random catalogue used for the stars was the same as that adopted for the uncontaminated mocks, as described in Sect.~\ref{sc:setup-angcorr-persistence}. For the persistence contaminants, we instead used a downsampled version of the same random catalogue, approximately 50 times larger than the persistence sample itself. This choice significantly reduced the computational cost while keeping shot-noise contributions negligible.

\begin{figure}[h!]
    \centering

    \includegraphics[width=\linewidth]{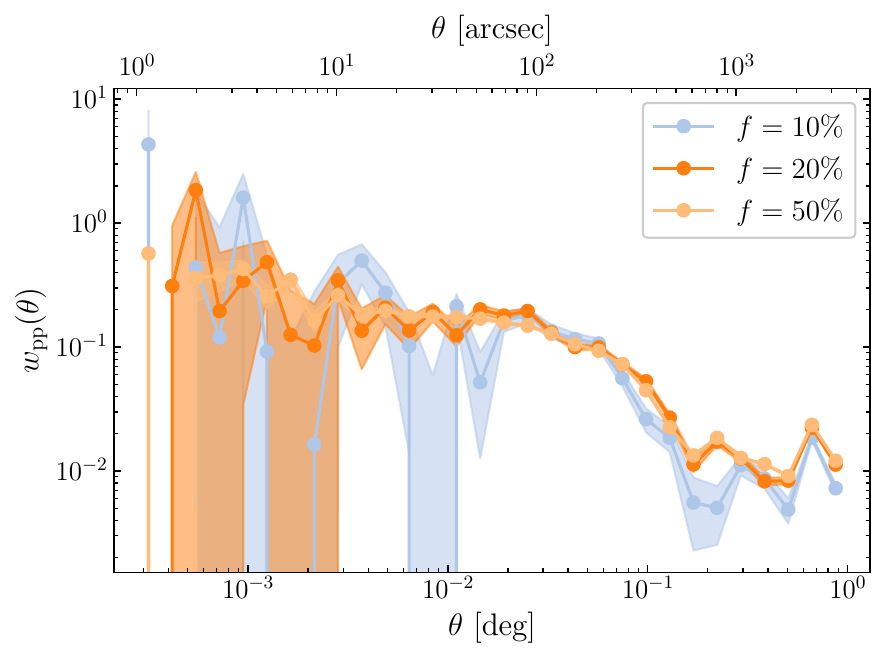}
    \hfill
    \includegraphics[width=\linewidth]{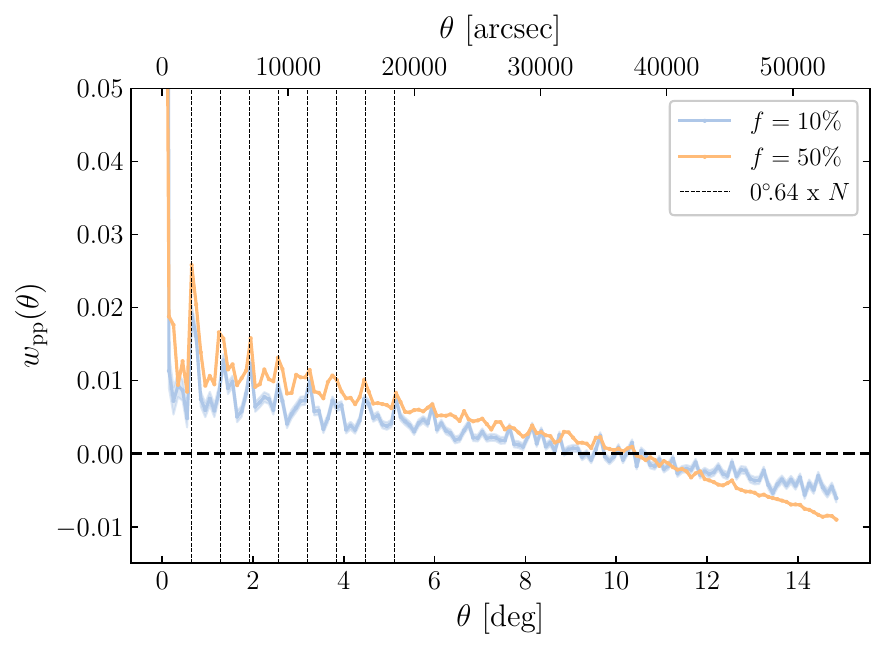}
    \hfill
    \includegraphics[width=\linewidth]{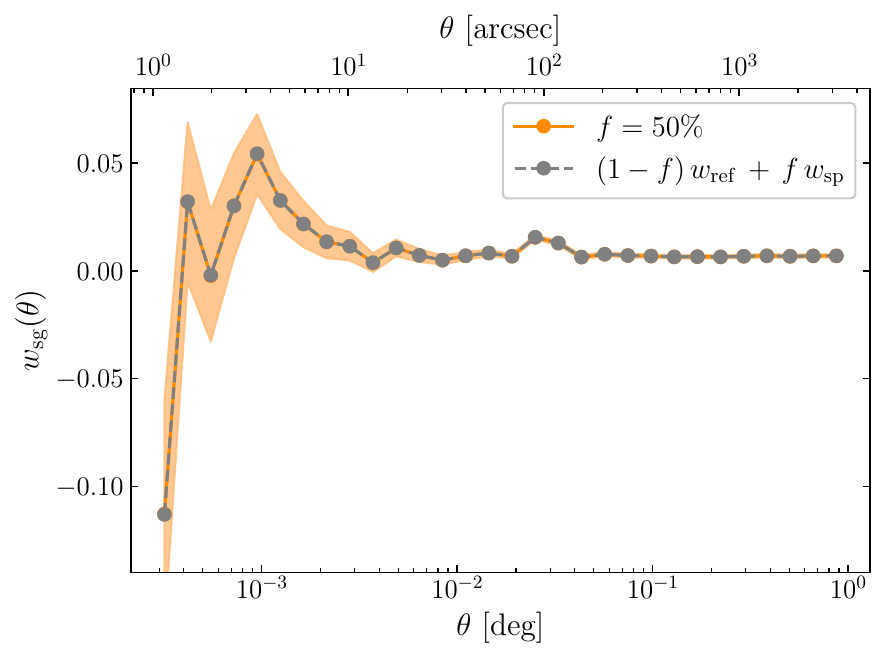}
    \caption{
    \textit{Top}: Persistence--persistence angular auto-correlation on sub-degree scales for $f=10\%$ and $f=50\%$. The shaded areas correspond to the scatter among the 100 mock measurements.
    \textit{Centre}: Large-scale persistence--persistence angular auto-correlation for $f=10\%$ and $f=50\%$. The coloured bands show the Poisson uncertainties.
    \textit{Bottom}: Star--galaxy angular correlation measured on mocks (orange curve) and the corresponding model for a contamination fraction $f=50\%$ (grey dashed line). The shaded area corresponds to the scatter among the 100 mock measurements.
    }
    \label{fig:persistence-overview}
\end{figure}

Figure~\ref{fig:persistence-overview}-top shows the persistence--persistence angular correlation on sub-degree scales for different contamination fractions. On these scales, the only difference is the smaller Poisson error in presence of more contaminants. Also, no strong periodic pattern is visible, because the signal results from the superposition of persistence replicas produced at different wavelengths and by different dither configurations. The situation changes at larger angular separations, shown in Fig.~\ref{fig:persistence-overview}-centre, where the periodic structure associated with the \Euclid dithering strategy becomes evident (vertical dashed lines). Persistence signals are replicated across the sky following the sequence of \Euclid pointings and, in particular, the characteristic angular scale of the field of view. This periodicity is responsible for the oscillatory behaviour observed in the galaxy--galaxy correlation in Fig.~\ref{fig:ls-angstat-f10f50} for $f=50\%$. The same pattern is also present for $f=10\%$, although strongly damped by the lower contamination fraction. More generally, the measured galaxy--galaxy correlation can be interpreted as the weighted combination of the true galaxy auto-correlation and the persistence--persistence auto-correlation.

A similar interpretation applies to the star--galaxy angular correlation, shown in Fig.~\ref{fig:persistence-overview}-bottom on sub-degree scales for a catalogue contaminated at the level $f = 50\%$ (orange points with $1\sigma$ uncertainty band). The measured correlation can be interpreted as the weighted combination of the correlation between stars and the uncontaminated galaxy sample, $w_{\mathrm{ref}}(\theta)$, expected to be consistent with zero, and the star--persistence cross-correlation, $w_{\mathrm{sp}}(\theta)$, such that
\begin{equation}
w(\theta) = (1 - f)\, w_{\mathrm{ref}}(\theta) + f\, w_{\mathrm{sp}}(\theta)\,.
\end{equation}
This expectation is confirmed in Fig.~\ref{fig:persistence-overview}-bottom, where the grey points show the function $w(\theta)$ reconstructed by combining the measured $w_{\mathrm{ref}}(\theta)$ and $w_{\mathrm{sp}}(\theta)$ from the mock catalogues.

\section{Impact of star globular clusters}
\label{app:globclusS1}

\begin{figure}[h]
    \centering
    \includegraphics[width=\linewidth]{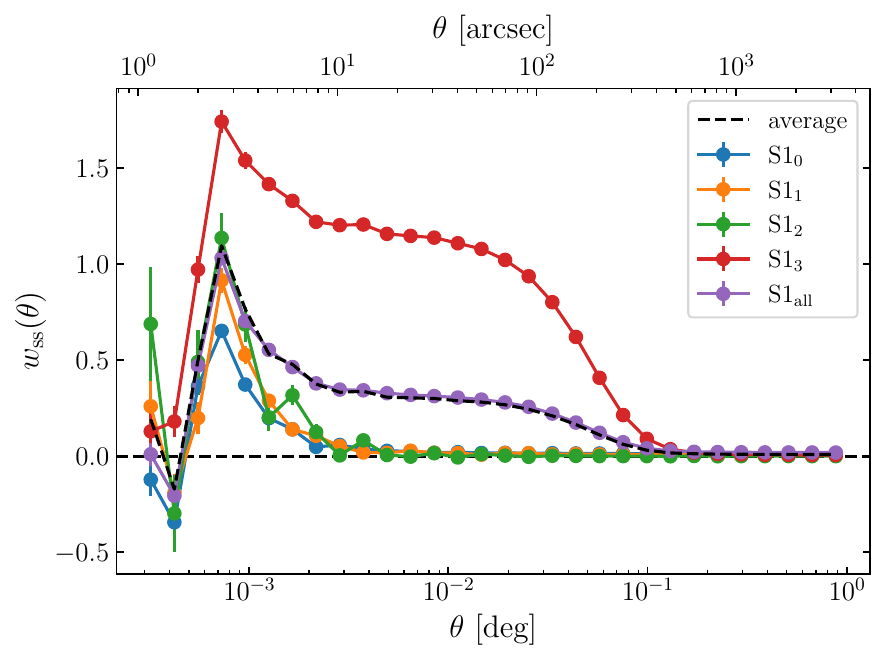}
    \caption{Star--star angular correlation in the S1 field. Different colours represent different sub-patches of S1, with the red line corresponding to the patch with the globular cluster inside. The average of the correlations in the different patches (black dashed line) matches the correlation computed on the whole S1 field (purple). The correlation in this plot was computed using \Gaia\ stars in order to enhance the intensity and shape of the signal compared to \twomass.}
    \label{fig:S1cluster4subpatchesGaia}
    \label{LastPage}
\end{figure}

Stars are not uniformly distributed either within or outside galaxies: prominent stellar structures such as groups and clusters are common.
The shape of the star--star correlation measured in the S1 field suggested the presence of such a structure, which we identified as the globular cluster NGC~1261. To verify this interpretation, we computed the star--star correlation in four different sub-regions of S1, only one of which contained the cluster. In all sub-patches excluding NGC~1261, the correlation shows the expected rapidly decreasing behaviour, whereas the sub-region containing the cluster reproduces the peculiar shape observed in the full S1 measurement (see Fig.~\ref{fig:S1cluster4subpatchesGaia}). 

\end{appendix}

\label{LastPage}
\end{document}